\documentclass{JHEP3}
\usepackage{amsmath,graphicx,epsfig,amssymb}

\def\em{\textrm{em}}
\def\lQ{\Lambda_{\rm QCD}} 
\newcommand{\onec}{1\!\!{\rm l}_c}
\newcommand{\nn}{\nonumber}
\newcommand{\be}{\begin{equation}}
\newcommand{\ee}{\end{equation}}
\newcommand{\bea}{\begin{eqnarray}}
\newcommand{\eea}{\end{eqnarray}}
\def\Dlr{-\frac{i}{2}\overleftrightarrow D}

\def\als{\alpha_{\rm s}}

\def\slashchar#1{\setbox0=\hbox{$#1$}\dimen0=\wd0\setbox1=\hbox{/}\dimen1=\wd1\ifdim\dimen0>\dimen1
                 \rlap{\hbox to \dimen0{\hfil/\hfil}} #1 \else \rlap{\hbox to
		   \dimen1{\hfil$#1$\hfil}} /  \fi}

\preprint{IFUM-854-FT}

\title{Electromagnetic quarkonium decays at order $v^7$}

\author{
Nora Brambilla, Emanuele Mereghetti\thanks{Present address: Physics Department, University of
Arizona, Tucson, AZ 85719.} ~and Antonio Vairo\\
Dipartimento di Fisica dell' Universit\'a di Milano and INFN, via Celoria 16,
20133 Milano, Italy\\
E-mail: \email{nora.brambilla@mi.infn.it}, \email{emanuele@physics.arizona.edu}, \email{antonio.vairo@mi.infn.it}
}

\abstract{
We compute S-wave and P-wave electromagnetic quarkonium decays at order $v^7$ 
in the heavy-quark velocity expansion. In the S-wave case, our calculation confirms 
and completes previous findings. In the P-wave case, our results are in disagreement 
with previous ones; in particular, we find that two matrix elements less are needed.
The cancellation of infrared singularities in the matching 
procedure is discussed.
}


\begin{document}

\section{Introduction}
In the last years, new measurements of heavy-quarkonium 
decay observables, mainly coming from BABAR, BELLE, BES, CLEO and
the Fermilab experiments have improved our knowledge of inclusive, electromagnetic and
several exclusive decay channels as well as of several electromagnetic and hadronic 
transition amplitudes. The new data and improved 
error analyses of the several correlated measurements have not 
only led to a sizeable reduction of the uncertainties but also, in some cases, 
to significant shifts in the central values \cite{Brambilla:2004wf}. 
Such data call for comparable accuracies in the theoretical determinations. 

The main mechanism of quarkonium decay into light particles is quark-antiquark annihilation. 
It happens at a scale which is twice the heavy-quark mass $M$.
Since this scale is perturbative, quark-antiquark annihilation may be described 
by an expansion in the strong-coupling constant $\als$. 
Experimentally this fact is reflected into the narrow widths 
of the quarkonia below the open flavor threshold. The bound-state dynamics, instead, 
is characterized by scales that are smaller than $M$. While we may not necessarily rely 
on an expansion in $\als$  to describe it, we may take advantage of the non-relativistic 
nature of the bound state and expand in the relative heavy-quark velocity $v$.
Decay-width formulas may be organized in a double expansion in $\als$, calculated at 
a large scale of the order of $M$, and in $v$. In the bottomonium system,  
typical reference values are $\als(M_b)\approx 0.2$, $v_b^2\approx 0.1$ and 
in the charmonium one, $\als(M_c)\approx 0.35$, $v_c^2\approx 0.3$.
We will call relativistic all corrections of order $v$ or smaller 
with respect to the leading decay-width expression. 

S-wave quarkonium decays into lepton pairs are known within a few percent
uncertainty. Theoretical accuracies of about 5\% in the charmonium 
case and of about 1\% in the bottomonium one require the calculation of ${O}(v^4,\als\,v^2,\als^2)$ corrections.  
Two photon decays are not so well known. The $\eta_c \to \gamma \gamma$ width
is known with an accuracy of 35\%, the $\chi_{c0}\to\gamma\gamma$ width 
with an accuracy of 20\% and the $\chi_{c2}\to\gamma\gamma$ width with an 
accuracy of 10\% \cite{Eidelman:2004wy}. Nevertheless, in the P-wave case,  
the improvement has been dramatic over the last few years: the errors
quoted in the 2000 edition of the Review of Particle Physics were 70\% 
for the  $\chi_{c0}\to\gamma\gamma$ case 
and 30\% for the $\chi_{c2}\to\gamma\gamma$ case \cite{Groom:2000in}. 
Theoretical accuracies matching the experimental ones require the calculation 
of ${O}(v^2,\als)$ corrections.  

In this work, we consider relativistic corrections of order $v^4$ and order $v^2$ 
to electromagnetic decays of S and P wave respectively.  Since the leading-order 
S-wave decay width is proportional to the square of the wave function in the origin, it 
is of order $v^3$. Since the leading order P-wave decay width is proportional 
to the square of the wave-function derivative in the origin, 
it is of order $v^5$. Therefore, both corrections of order $v^4$ to S-wave decays and 
of order $v^2$ to P-wave decays provide  decay widths at order $v^7$.

In the S-wave case, corrections of order $v^2$ and $v^4$ were first considered 
in \cite{Bodwin:1994jh} and \cite{Bodwin:2002hg} respectively. We agree with their results
if we use their power counting, but we find three new contributions using our
different counting. Moreover, we resolve an ambiguity in the matching coefficients of the operators contributing
at order $v^4$. In the P-wave case, corrections of order $v^2$ were first calculated 
in \cite{Ma:2002ev}. Our results disagree with those in \cite{Ma:2002ev}. We find that we need two 
operators less to describe the decay widths at that order. 

This work is partially based on \cite{Emanuele}. We refer to it for some detailed derivations.

The paper is organized in the following way. In Sec. \ref{basics}, we set up
the formalism, discuss the power counting and introduce our basis of operators. In Sec. \ref{matching}, we
perform the matching. We also discuss the cancellation of the infrared 
singularities in the matching of the octet operators. In Sec. \ref{conclusions},
we conclude by discussing applications and further developments of this work.
In Appendix \ref{AppA} and \ref{AppB}, we list all the operators and the  matching 
coefficients that have been employed throughout the paper.

\section{Decay widths in NRQCD}
\label{basics}
\subsection{NRQCD}
In the effective field theory (EFT) language of non-relativistic QCD 
(NRQCD), the annihilation of quarkonium is described 
by four-fermion operators \cite{Bodwin:1994jh}.
The decay width factorizes in a high-energy contribution  encoded 
in the imaginary parts of the four-fermion matching coefficients 
and a low-energy contribution encoded in the matrix elements of the
four-fermion operators evaluated on the heavy-quarkonium state.
The NRQCD factorization formula for the electromagnetic
decay width of a quarkonium $H$ is 
\be
\Gamma(H\to {\em}) = 2 \sum_n 
\frac{{\rm Im} \, c_{\em}^{(n)}}{M^{d_n - 4}} 
\langle H |\mathcal O^{(n)}_{\textrm{4-f}\,\em} |H \rangle.
\label{factannem}  
\ee
In this work, $H\to {\em}$ stands either for $H\to\gamma\gamma$ or for $H\to e^+e^-$.
$| H\rangle$ is a dimension $-3/2$ normalized eigenstate  
of the NRQCD Hamiltonian with the quantum numbers of the quarkonium $H$.
$\mathcal O^{(n)}_{\textrm{4-f}\,\em}$ is a generic four-fermion operator 
of dimension $d_n$. The ``${\em}$'' subscript means that we have singled out 
the electromagnetic component by projecting onto the QCD vacuum state $ |0\rangle$. 
The general form of $\mathcal O^{(n)}_{\textrm{4-f}\,\em}$ is 
$\psi^\dagger (\cdots) \chi \, |0\rangle \, \langle 0|\, \chi^\dagger (\cdots) \psi$, 
were $\psi$ is the Pauli spinor that annihilates a
quark and $\chi$ is the one that creates an antiquark.
The operators $(\cdots)$ may transform as singlets or octets under SU(3) gauge transformations.
A list of four-fermion operators is provided in Appendix \ref{AppA}.
The coefficients $c_{\em}^{(n)}$ encode the high-energy contributions 
to the electromagnetic annihilation processes, which have been integrated out from the EFT. 
At variance with \cite{Bodwin:1994jh}, we will not label differently 
coefficients that stem from decays into $\gamma\gamma$ or $e^+e^-$.
The coefficients  are calculated in perturbation theory by 
matching Green functions or physical amplitudes in QCD 
and NRQCD. It is the purpose of this work to calculate 
${\rm Im} \,c_{\em}^{(n)}$ for operators up to dimension 10.

\subsection{Power Counting}
\label{powcou}
In the factorization formula \eqref{factannem}, the matching coefficients 
$c_{\em}^{(n)}$ are series in $\als$ while the matrix elements 
$\langle H |\mathcal O^{(n)}_{\textrm{4-f}\,\em} |H \rangle$
are series in $v$ and are, in general, 
non-perturbative objects.   
It is not possible to attribute a definite power counting 
to the matrix elements because of the several contributing energy scales, 
as it is typical in  a non-relativistic system: $Mv$, $Mv^2$, ... .
Whatever power counting one assumes, as long as $v\ll 1$,  
operators of higher dimensionality are suppressed by powers of $v$.

In the following, we  will assume $Mv$ to be of the same order as the 
typical hadronic scale $\lQ$ and adopt the following counting. 
Matrix elements of the type $\langle H' |\mathcal O |H \rangle$, 
where $\mathcal O |H \rangle$ and $|H' \rangle$ have 
the same quantum numbers and color transformation properties
in the dominant Fock state, scale (at leading order) like $(Mv)^{d-3}$, $d$ being 
the dimension of the  operator $\mathcal O$.
If $\mathcal O |H \rangle$ and $|H' \rangle$ do not have the dominant Fock
state with the same quantum numbers, then the matrix element singles out a 
component of the quarkonium Fock state that is suppressed.
The amount of suppression depends on the power counting and 
on the quantum numbers. Let us consider a component in $\vert H' \rangle$ 
that shows up as a first-order correction to a singlet quarkonium state with orbital angular 
momentum $L$, spin $S$ and total angular momentum $J$. The correction is  
induced by an operator $\mathcal F$ of dimension $d_{\mathcal F}$: 
\be
\sum_{S',L',J'}  |^{2S'+1}L'_{J'} \rangle
\frac{\langle ^{2S'+1}L'_{J'} | \displaystyle 
\frac{1}{M^{d_{\mathcal F}-4}} \, {\mathcal F} \,|^{2S+1}L_J\rangle}{\displaystyle  E-E'}.
\label{eq:1def}
\ee
This gives a nonvanishing contribution
to the matrix element  $\langle H' |\mathcal O |H \rangle$
if for some $S'$, $L'$ and $J'$ the matrix element in \eqref{eq:1def} does not vanish and 
if there is a Fock state of $\mathcal O |H \rangle$ with the same quantum
numbers as $|^{2S'+1}L'_{J'} \rangle$. In case this is the dominant Fock state 
of $\mathcal O |H \rangle$, the suppression factor with respect 
to the naive scaling, $(Mv)^{d-3}$, is $M v^{d_{\mathcal F}-3}/(E-E')$.

As an example, we consider the dimension 5 operator 
$\mathcal F = \psi^\dagger i \, t^a \, g\vec A^a \cdot \vec{\partial} \psi + \textrm{c.c.}$, 
where $t^a$ are the color matrices in the fundamental representation and 
c.c. stands for ``charge conjugated'' ($\psi \rightarrow i \sigma ^2 \chi^*$, 
$t^a \, A^a_\mu \rightarrow - (t^a \,A^a_\mu)^T$).
Inserted in \eqref{eq:1def}, it projects onto an octet state 
with quantum numbers $L' = L \pm 1$ and $S'= S$. 
Color octet quark-antiquark states may appear in the quarkonium spectrum 
combined with gluons to form hybrids. Gluonic excitations of this kind are expected to 
develop a mass gap of order  $\lQ$, hence $E-E'\sim \lQ \sim Mv$. 
As a consequence, in a quarkonium state the octet component with quantum numbers $S$ and $L\pm 1$  
is suppressed by $v$ with respect to the singlet component with quantum numbers $S$ and $L$.
Note that to generate a D-wave component ($L =2$) from a S-wave state we need 
two operator insertions and second-order perturbation theory. 
Hence, this is suppressed by $v^2$ with respect to the S-wave component.

A similar counting holds if we consider the dimension 5 operator 
$\mathcal F = \psi^\dagger \, t^a \, g\vec B^a \cdot \vec{\sigma}\psi + \textrm{c.c.}$.
Inserted in \eqref{eq:1def}, it projects onto an octet state 
with quantum numbers $L' = L$ and $S'= S\pm 1$. 
Thus, also  the octet component with quantum numbers $S\pm 1$ and $L$ 
is suppressed by $v$ with respect to the singlet component with quantum numbers 
$S$ and $L$.

The power counting adopted here is the most conservative one in the framework 
of NRQCD. This is the reason for our choice. The price we pay is that some
observables will depend on more matrix elements than they would in a different 
counting. An example of alternative power counting, which seems better suited 
for the situation $Mv^2 \sim \lQ$, is provided by Ref. \cite{Bodwin:1994jh}.
For a critical review and a discussion on the different power countings 
we refer to \cite{Brambilla:2004jw} and references therein.

\subsection{Four-fermion operators}
The four-fermion sector of the NRQCD Lagrangian contains
all four-fermion operators invariant under gauge transformations,  
rotations, translations, charge conjugation, parity and time 
inversion. They may be classified according to their dimensionality and 
color content. Moreover, it is useful to decompose them in terms of 
irreducible spherical tensors. Some of the operators are redundant 
because they may be expressed in terms of others by means of 
field redefinitions. Finally, the power counting introduced in the 
previous section sets the relevance of the different operators 
in the decay width formulas. We will address all these issues in the following.

\subsubsection{Operators from dimension 6 to 10}
The NRQCD operators may be organized according to their dimensionality.
Four-fermion operators of dimension 6 were considered in
\cite{Bodwin:1994jh}. For dimensional reasons, they cannot depend 
on the gluon fields. The only allowed color structures are $\onec \otimes \onec$
and $t^a \otimes t^a$. However, since we are considering electromagnetic
decays whose final state is the QCD vacuum, the $t^a \otimes t^a$ 
structure is forbidden. 

Parity conservation forbids four-fermion operators of dimension 7.
Four-fermion operators of dimension 8 may be built 
with  two covariant derivatives \cite{Bodwin:1994jh}, or with  a chromomagnetic 
field, like, for instance,
\be
\psi^{\dag} g\vec B  \cdot \vec{\sigma}\chi |0\rangle \langle 0|
\chi^{\dag} \psi + \textrm{H.c.},
\label{SB1}
\ee
where H.c. stands for Hermitian conjugated. 
The operator describes a (spin-flipping) octet to singlet $Q \bar Q$ pair transitions. 
The same operator with a chromoelectric field instead of the 
chromomagnetic one is forbidden by parity conservation.

Operators of dimension 9 may involve a covariant
derivative and a chromoelectric field, like, for instance,
\be
\psi^{\dag} \overleftrightarrow D \cdot \vec{\sigma} \chi |0\rangle 
\langle 0| \chi^{\dag}  g\vec E \cdot \vec{\sigma} \psi 
+ \textrm{H.c.}.
\label{SESD}
\ee
The same operator with a chromomagnetic field instead of the 
chromoelectric one is forbidden by parity conservation.

Dimension 10 operators may involve either four
covariant derivatives, or two covariant derivatives and a
chromomagnetic field, like, for instance, 
\be
\psi^{\dag} g\vec B  \cdot \overleftrightarrow {D}\chi |0\rangle
\langle 0| \chi^{\dag} \overleftrightarrow D \cdot \vec\sigma\psi +
\textrm{H.c.},
\ee
or two gluon fields.

\subsubsection{Irreducible spherical tensors}
All the tensorial structures that are consistent with the  discrete symmetries
and rotational invariance are allowed. So, for instance, besides 
Eq. \eqref{SB1} also the following operator is possible
\be
\psi^{\dag}  g\vec B  \chi |0\rangle \langle 0| \chi^{\dag}
\vec{\sigma} \psi + \textrm{H.c.}.
\label{SB1bis}
\ee 

It is useful to decompose Cartesian tensors into irreducible spherical tensors.
This allows a classification of the operators in terms
of angular-momentum quantum numbers: $^{2S+1}L_J$, where
$S$ is the spin, $L$ the orbital and $J$ the total angular momentum. 

According to $\mathbf{3}\otimes \mathbf{3} = \mathbf{5}  \oplus \mathbf{3} \oplus \mathbf{1}$, 
a tensor made of two vectors $\vec{A}$ and $\vec{B}$ may be decomposed 
as
\be
A^iB^j = A^{(i}B^{j)}
+ \frac{A^iB^j-A^jB^i}{2}
+ \frac{\delta^{ij}}{3} \vec{A}\cdot\vec{B},
\label{dec2}
\ee
where 
\be
A^{(i}B^{j)}=\frac{A^iB^j+A^jB^i}{2}-\frac{\delta^{ij}}{3}\vec A
\cdot \vec B,
\label{eq:B.b}
\ee
is a symmetric and traceless tensor belonging to the
representation $L=2$ of the rotational group, 
and the last two terms of Eq. \eqref{dec2} are tensors transforming respectively 
as $L=1$ and $L=0$ spherical harmonics.
So, for instance,  besides Eq. \eqref{SESD} also the following operator has to be considered
\be
\psi^{\dag}  \overleftrightarrow D^{(i}  \sigma^{j)}  \chi |0\rangle 
\langle 0| \chi^{\dag} g\vec E^{(i}  \sigma^{j)} \psi 
+ \textrm{H.c.}.
\label{SESDbis}
\ee
Note that a symmetric tensor $S^{ij}$ may be decomposed, 
according to  $\mathbf{6}= \mathbf{5} \oplus \mathbf{1}$, into
\be
S^{ij} = S^{(ij)} + \frac{\delta^{ij}}{3} S^{kk},
\ee
where $S^{(ij)}$ is a traceless symmetric tensor.

Therefore, from $(\mathbf{5} \oplus \mathbf{1}) \otimes \mathbf{3}
= \mathbf{7} \oplus \mathbf{5} \oplus \mathbf{3}\oplus \mathbf{3}$, 
a tensor made of a symmetric tensor $S^{ij}$ and a vector $\vec{A}$ may be
decomposed into 
\be
\begin{gathered}
S^{ij}A^{k} =
S^{((ij)}A^{k)} 
+\frac{\epsilon^{ikl}\delta^{jm}+\epsilon^{jkl}\delta^{im}}{3}
\left(\epsilon^{lnp}S^{(mn)}A^p+\frac{\epsilon^{lmp}S^{(np)}A^n}{2}\right)
\\
-
\frac{3}{10}\left(\frac{2}{3}\delta^{ij}\delta^{lk}- \delta^{kj}\delta^{il}- \delta^{ki}\delta^{jl}\right)
S^{(ml)}A^m
+ \frac{\delta^{ij}}{3}S^{mm}A^k,
\end{gathered}
\label{eq:B.c}
\ee
were $S^{((ij)}A^{k)}$ stands for a tensor symmetric with respect to
the indices  $i$, $j$ and $k$, and such that all partial traces 
(obtained contracting two of the three indices) are equal to zero:
\be
S^{((ij)}A^{k)} =
\frac{1}{3}\left(S^{(ij)}A^k+S^{(ik)}A^j+S^{(jk)}A^i\right)
-\frac{2}{15}\left(\delta^{ij}\delta^{lk}+\delta^{ik}\delta^{lj}+\delta^{jk}\delta^{li}
\right)S^{(ml)}A^m.
\label{eq:B.d}
\ee
The second term in Eq. \eqref{eq:B.c} transforms like a 
$L=2$ spherical harmonics. Note that $\epsilon^{lnp}S^{(mn)}A^p+{\epsilon^{lmp}S^{(np)}A^n}/{2}$
is symmetric in the indices $l$ and $m$ and traceless.
The third and fourth terms in Eq. \eqref{eq:B.c} transform like vectors.

As an application, we may use Eq. \eqref{eq:B.c} to decompose the tensor 
\be
\left(-\frac{i}{2}\overleftrightarrow D^{i} \right)
\left(-\frac{i}{2}\overleftrightarrow D^{j} \right)
\left(-\frac{i}{2}\overleftrightarrow D^{k} \right)
\ee
into irreducible spherical tensors. If we treat the covariant derivatives as ordinary ones, 
the tensor, being completely symmetric, has 10 independent components. 
According to $\mathbf{10} = \mathbf{7} \oplus \mathbf{3}$, it may be decomposed into a $L=3$ tensor
\be 
{\bf T}^{(3)}_{ijk} = 
\left(-\frac{i}{2}\right)^2\overleftrightarrow D^{(i}\overleftrightarrow D^{j)}
\left(-\frac{i}{2}\overleftrightarrow D^k\right)
 - \frac{2}{5}
\left(-\frac{i}{2}\right)\overleftrightarrow D^{(i} \delta^{j)k}
\left(- \frac{i}{2} \overleftrightarrow  D \right)^2,
\label{T3tensor}
\ee
where $A^{(i}\delta^{j)k}= A^i\delta^{jk}/2 +A^j\delta^{ik}/2 - A^k \delta^{ij}/3$, 
and a tensor that transforms like a $L=1$ representation of the rotational group:
\bea
{\bf T}^{(1)}_{ijk} = 
\frac{1}{5}\left(\delta^{ij}\delta^{lk} + \delta^{kj}\delta^{il} + \delta^{ki}\delta^{lj}\right)
\left(- \frac{i}{2} \overleftrightarrow D^{l} \right)
\left(-\frac{i}{2} \overleftrightarrow D \right)^2.
\eea
From these we may construct the four-fermion operators listed in Eq. \eqref{eq:B.5}:
\bea
\hspace{-10mm}
\mathcal P_{\textrm{em}}(^3P_0) &=& 
\frac{1}{2} \psi^{\dag} {\bf T}^{(1)}_{ijk} \sigma^k \chi |0 \rangle 
\langle 0|\chi^{\dag}\,\frac{\delta^{ij}}{3}\left(-\frac{i}{2}\overleftrightarrow D \cdot \vec{\sigma}\right)\psi 
+ \textrm{H.c.},
\\
\hspace{-10mm}
\mathcal P_{\textrm{em}}(^3P_2) &=& 
\frac{5}{4} \psi^{\dag} \left(
\frac{{\bf T}^{(1)}_{ijk} + {\bf T}^{(1)}_{jik}}{2} - \frac{\delta^{ij}}{3}{\bf T}^{(1)}_{llk}
\right) \sigma^k \chi |0 \rangle 
\langle 0| \chi^{\dag}\left(-\frac{i}{2}\overleftrightarrow D^{(i}\sigma^{j)}\right)\psi 
+ \textrm{H.c.},
\\
\hspace{-10mm}
\mathcal P_{\textrm{em}}(^3P_2,^3F_2) &=& 
\frac{1}{2} \psi^{\dag} {\bf T}^{(3)}_{ijk} \sigma^k \chi |0 \rangle 
\langle 0| \chi^{\dag}\left(-\frac{i}{2}\overleftrightarrow D^{(i}\sigma^{j)}\right)\psi 
+ \textrm{H.c.}.
\label{T3PF}
\eea

\subsubsection{Field redefinitions}
\label{secFR}
Operators may be redundant, in the sense that they  may be traded for others by means 
of suitable field redefinitions. Let us consider the following cases.

{\it (i)} First, we consider the field redefinitions ($a$ is a free parameter)
\be
\left\{
\begin{gathered}
\psi \rightarrow \psi + \frac{a}{M^5} \left[ \left(\Dlr\right)^2,\chi |0\rangle \langle 0|\chi^\dagger \right]\psi
\\
\chi \rightarrow \chi - \frac{a}{M^5} \left[ \left(\Dlr\right)^2,\psi |0\rangle \langle 0|\psi^\dagger \right]\chi
\end{gathered}
\right.,
\label{fieldred1}
\ee
which induce the following transformations:
\bea
&& \hspace{-8mm}
\psi^\dagger \, i D_0 \, \psi + \chi^\dagger \, iD_0 \, \chi 
\rightarrow \psi^\dagger \, i D_0 \, \psi + \chi^\dagger \, iD_0 \, \chi 
- \frac{a}{M^5}\, \mathcal T_{8\,\textrm{em}}(^1S_0),
\\
&& \hspace{-8mm}
\psi^\dagger \, \frac{{\vec{D}}^2}{2M} \, \psi - \chi^\dagger \, \frac{{\vec{D}}^2}{2M}  \, \chi 
\rightarrow \psi^\dagger \, \frac{{\vec{D}}^2}{2M}  \, \psi - \chi^\dagger \, \frac{{\vec{D}}^2}{2M}  \, \chi 
+ 2 \frac{a}{M^6}\,\left( \mathcal{Q'}_{\textrm{em}}(^1S_0) - \mathcal{Q''}_{\textrm{em}}(^1S_0) \right),
\label{FRkin1}
\eea
where the operator $\mathcal T_{8\,\textrm{em}}(^1S_0)$ has been defined in Eq. \eqref{eq:B.9}
and the operators $\mathcal{Q'}_{\textrm{em}}(^1S_0)$ and $\mathcal{Q''}_{\textrm{em}}(^1S_0)$
in Eq. \eqref{eq:B.4}. In Eq. \eqref{FRkin1} we have neglected operators proportional 
to the center of mass momentum and octet operators, whose matrix elements are subleading in any power counting.
Clearly, for a suitable choice of the parameter $a$ we may eliminate the operator $\mathcal T_{8\,\textrm{em}}(^1S_0)$ 
from our basis of operators in exchange for a redefinition of the matching coefficients 
of $\mathcal{Q'}_{\textrm{em}}(^1S_0)$ and $\mathcal{Q''}_{\textrm{em}}(^1S_0)$:
$h'_{\textrm{em}}(^1S_0) \to  h'_{\textrm{em}}(^1S_0) + 2 a$ and $h''_{\textrm{em}}(^1S_0) \to  h''_{\textrm{em}}(^1S_0) - 2 a$.
Note that the sum $h'_{\textrm{em}}(^1S_0) + h''_{\textrm{em}}(^1S_0)$ is invariant under the field 
redefinition.

{\it (ii)} Second, we consider, the field redefinitions
\be
\left\{
\begin{gathered}
\psi \;\;\lower5pt\vbox{\hbox{\rlap{\tiny $J$}\lower-5pt\vbox{\hbox{$\!
       \rightarrow$}}}}\;\;
\psi + \frac{a}{M^5} \, {\bf T}_{ijlk}^{(J)} \, \sigma^l 
\left[ \left(\Dlr^i\right)\left(\Dlr^j\right),\chi |0\rangle \langle 0|\chi^\dagger \right]\sigma^k\psi
\\
\chi \;\;\lower5pt\vbox{\hbox{\rlap{\tiny $J$}\lower-5pt\vbox{\hbox{$\!
       \rightarrow$}}}}\;\;
\chi - \frac{a}{M^5} \, {\bf T}_{ijlk}^{(J)} \, \sigma^l
\left[ \left(\Dlr^j\right)\left(\Dlr^i\right) ,\psi |0\rangle \langle 0|\psi^\dagger \right]\sigma^k\chi
\end{gathered}
\right.,
\label{fieldred2}
\ee
where 
\bea
&& {\bf T}_{ijlk}^{(0)} = \frac{\delta^{ij}\delta^{lk}}{3},
\\
&& {\bf T}_{ijlk}^{(1)} = \frac{\epsilon_{ijn}\epsilon_{kln}}{2},
\\
&& {\bf T}_{ijlk}^{(2)} = \frac{\delta^{il}\delta^{jk} + \delta^{jl}\delta^{ik}}{2} -\frac{\delta^{ij}\delta^{lk}}{3}.
\eea
They induce the following transformations:
\bea
&& 
\psi^\dagger \, i D_0 \, \psi + \chi^\dagger \, iD_0 \, \chi 
\;\;\lower5pt\vbox{\hbox{\rlap{\tiny $J$}\lower-5pt\vbox{\hbox{$\!
       \rightarrow$}}}}\;\;
\psi^\dagger \, i D_0 \, \psi + \chi^\dagger \, iD_0 \, \chi 
- \frac{a}{M^5}\, \mathcal T_{8\,\textrm{em}}^{(J)}(^3S_1),
\\
&& 
\psi^\dagger \, \frac{{\vec{D}}^2}{2M} \, \psi - \chi^\dagger \, \frac{{\vec{D}}^2}{2M}  \, \chi 
\;\;\lower5pt\vbox{\hbox{\rlap{\tiny $\!\!J\!=\!0$}\lower-5pt\vbox{\hbox{$\!
       \longrightarrow$}}}}\;\;
\psi^\dagger \, \frac{{\vec{D}}^2}{2M}  \, \psi - \chi^\dagger \, \frac{{\vec{D}}^2}{2M}  \, \chi 
\nn\\
&& \hspace{50mm}
+ 2 \frac{a}{M^6}
\,\left( \mathcal{Q'}_{\textrm{em}}(^3S_1) - \mathcal{Q''}_{\textrm{em}}(^3S_1) \right),
\label{FRkin2}
\\
&& 
\psi^\dagger \, \frac{{\vec{D}}^2}{2M} \, \psi - \chi^\dagger \, \frac{{\vec{D}}^2}{2M}  \, \chi 
\;\;\lower5pt\vbox{\hbox{\rlap{\tiny $\!\!J\!=\!1$}\lower-5pt\vbox{\hbox{$\!
       \longrightarrow$}}}}\;\;
\psi^\dagger \, \frac{{\vec{D}}^2}{2M}  \, \psi - \chi^\dagger \, \frac{{\vec{D}}^2}{2M}  \, \chi 
\nn\\
&& \hspace{50mm}
+ \frac{1}{M^6} \times \textrm{(octet operators)},
\label{FRkin3}
\\
&& 
\psi^\dagger \, \frac{{\vec{D}}^2}{2M} \, \psi - \chi^\dagger \, \frac{{\vec{D}}^2}{2M}  \, \chi 
\;\;\lower5pt\vbox{\hbox{\rlap{\tiny $\!\!J\!=\!2$}\lower-5pt\vbox{\hbox{$\!
       \longrightarrow$}}}}\;\;
\psi^\dagger \, \frac{{\vec{D}}^2}{2M}  \, \psi - \chi^\dagger \, \frac{{\vec{D}}^2}{2M}  \, \chi 
\nn\\
&& \hspace{50mm}
+ 2 \frac{a}{M^6}
\,\left( \mathcal{Q'}_{\textrm{em}}(^3S_1,^3D_1) - \mathcal{Q''}_{\textrm{em}}(^3S_1,^3D_1) \right),
\label{FRkin4}
\eea
where the operators $\mathcal T_{8\,\textrm{em}}^{(J)}(^3S_1)$ have been defined in Eq. \eqref{eq:B.9}
and the operators $\mathcal{Q'}_{\textrm{em}}(^3S_1)$, $\mathcal{Q''}_{\textrm{em}}(^3S_1)$, 
$\mathcal{Q'}_{\textrm{em}}(^3S_1,^3D_1)$ and $\mathcal{Q''}_{\textrm{em}}(^3S_1,^3D_1)$
in Eq. \eqref{eq:B.4}. In Eqs. \eqref{FRkin2}-\eqref{FRkin4} we have neglected
operators proportional to the center of mass momentum and octet operators.
Again, for a suitable choice of the parameter $a$ we may eliminate the operators $\mathcal T_{8\,\textrm{em}}^{(J)}(^3S_1)$ 
from our basis of operators in exchange for a redefinition of the matching coefficients 
of $\mathcal{Q'}_{\textrm{em}}(^3S_1)$, $\mathcal{Q''}_{\textrm{em}}(^3S_1)$, 
$\mathcal{Q'}_{\textrm{em}}(^3S_1,^3D_1)$ and $\mathcal{Q''}_{\textrm{em}}(^3S_1,^3D_1)$.
Note that $h'_{\textrm{em}}(^3S_1) + h''_{\textrm{em}}(^3S_1)$ and 
$h'_{\textrm{em}}(^3S_1,^3D_1) + h''_{\textrm{em}}(^3S_1,^3D_1)$
are invariant under the field redefinitions.

In \cite{Bodwin:2002hg}, it was pointed out that the operators $\mathcal T_{8\,\textrm{em}}(^1S_0)$ and 
$\mathcal T_{8\,\textrm{em}}(^3S_1)$ may be eliminated in favor of $\mathcal{Q'}_{\textrm{em}}(^1S_0)$ and 
$\mathcal{Q''}_{\textrm{em}}(^1S_0)$, and  $\mathcal{Q'}_{\textrm{em}}(^3S_1)$ and 
$\mathcal{Q''}_{\textrm{em}}(^3S_1)$ through the use of the equations of motion.
Our argument, which uses field redefinitions instead of equations of motion, is equivalent 
\cite{Kilian:1994mg}.

{\it (iii)} Finally, we may wonder if there are field redefinitions similar to \eqref{fieldred1} and 
\eqref{fieldred2} that allow to eliminate from our basis also the operators 
$\mathcal T_{8\,\textrm{em}}(^3P_J)$ defined in Eq. \eqref{eq:B.9}.
Indeed, such redefinitions exist and are
\be
\left\{
\begin{gathered}
\psi \;\;\lower5pt\vbox{\hbox{\rlap{\tiny $J$}\lower-5pt\vbox{\hbox{$\!
       \rightarrow$}}}}\;\;
\psi + \frac{a}{M^5} \, {\bf T}_{ijlk}^{(J)} \, \sigma^i 
\left(\Dlr^j\right)\chi|0\rangle \langle 0|\chi^\dagger \sigma^l \left(\Dlr^k\right)\psi
\\
\chi \;\;\lower5pt\vbox{\hbox{\rlap{\tiny $J$}\lower-5pt\vbox{\hbox{$\!
       \rightarrow$}}}}\;\;
\chi - \frac{a}{M^5} \, {\bf T}_{ijlk}^{(J)} \, \sigma^i
\left(\Dlr^j\right) \psi|0\rangle \langle 0|\psi^\dagger \sigma^l \left(\Dlr^k\right) \chi
\end{gathered}
\right. .
\label{fieldred3}
\ee
However, in this case, they induce the following transformations
\bea
&& 
\psi^\dagger \, i D_0 \, \psi + \chi^\dagger \, iD_0 \, \chi 
\;\;\lower5pt\vbox{\hbox{\rlap{\tiny $J$}\lower-5pt\vbox{\hbox{$\!
       \rightarrow$}}}}\;\;
\psi^\dagger \, i D_0 \, \psi + \chi^\dagger \, iD_0 \, \chi 
- \frac{a}{M^5}\mathcal T_{8\,\textrm{em}}(^3P_J)  
\nn\\
&& \hspace{25mm}
- 2 \frac{a}{M^5}
{\bf T}_{ijlk}^{(J)}\, i\partial_0 \left(\psi^\dagger\sigma^i \left(\Dlr^j\right) \chi\right)
|0\rangle \langle 0| \chi^\dagger \sigma^l \left(\Dlr^k\right) \psi,
\\
&& 
\psi^\dagger \, \frac{{\vec{D}}^2}{2M} \, \psi - \chi^\dagger \, \frac{{\vec{D}}^2}{2M}  \, \chi 
\;\;\lower5pt\vbox{\hbox{\rlap{\tiny $J$}\lower-5pt\vbox{\hbox{$\!
       \rightarrow$}}}}\;\;
\psi^\dagger \, \frac{{\vec{D}}^2}{2M}  \, \psi - \chi^\dagger \, \frac{{\vec{D}}^2}{2M}  \, \chi 
- 2 \frac{a}{M^6} \mathcal{P}_{\textrm{em}}(^3P_J),
\label{FRkin5}
\eea
where in Eq. \eqref{FRkin5} we have neglected operators proportional 
to the center of mass momentum and octet operators.
We see that the field redefinitions \eqref{fieldred3} may be useful to trade the 
operators  $\mathcal T_{8\,\textrm{em}}(^3P_J)$ for 
${\bf T}_{ijlk}^{(J)}\,$  $i\partial_0 \left(\psi^\dagger\sigma^i \left(\Dlr^j\right) \chi\right) |0\rangle$ 
$\langle 0| \chi^\dagger \sigma^l \left(\Dlr^k\right) \psi$, but, differently from the previous 
cases, they do not reduce the number of operators.
We will keep $\mathcal T_{8\,\textrm{em}}(^3P_J)$ in our basis of operators.

\subsubsection{Power counting of the four-fermion operators}
\label{secpowfour}
From the rules of Sec. \ref{powcou}, it follows that 
\be
\langle H(^{2S+1}L_J) | \frac{1}{M^{d-4}} \, {\mathcal O}_\textrm{em}(^{2S+1}L_J)| H(^{2S+1}L_J) \rangle
\sim M v^{d-3},
\ee
where $| H(^{2S+1}L_J) \rangle$ stands for a quarkonium state whose
dominant Fock-space component is a $Q \bar Q$ pair with quantum numbers $S$,
$L$ and $J$, ${\mathcal O}_\textrm{em}(^{2S+1}L_J)$ is a singlet 
four-fermion operator that acts on the  $Q \bar Q$ pair like a spin $S$, 
orbital angular momentum $L$ and total angular momentum $J$ tensor and $d$ 
is the dimension of the operator.

Concerning the power counting of the octet matrix elements, 
first, we consider the matrix elements of the dimension 8 operators defined 
in Eq. \eqref{eq:B.8}:
\be
\langle H(^1S_0) |  \frac{1}{M^{4}}\, \mathcal S_{8\,\em}(^1S_0) | H(^1S_0) \rangle,
\ee
and 
\be
\langle H(^3S_1) |  \frac{1}{M^{4}}\, \mathcal S_{8\,\em}(^3S_1) | H(^3S_1) \rangle.
\ee
The operator $\mathcal S_{8\,\em}(^1S_0)$ 
destroys a singlet $Q \bar Q$ pair with quantum number $^1S_0$ and creates
an octet $Q \bar Q$ pair with quantum numbers $^3S_1$ and a gluon (and viceversa),
the operator $\mathcal S_{8\,\em}(^3S_1)$ 
destroys a singlet $Q \bar Q$ pair with quantum number $^3S_1$ and creates
an octet $Q \bar Q$ pair with quantum numbers $^1S_0$ and a gluon (and viceversa).
Following the power counting  of Sec. \ref{powcou}  
the chromomagnetic field scales like $(Mv)^2$, moreover the octet Fock-space component
is suppressed by $v$. Hence, both matrix elements scale like $M v^6$.

Equations \eqref{eq:B.9} define octet operators of dimension 9.
The operator $\mathcal T^{(1)\prime}_{8\,\textrm{em}}(^3S_1)$ 
destroys a singlet $Q \bar Q$ pair with quantum number $^3S_1$ and creates
an octet $Q \bar Q$ pair with quantum numbers $^3P_1$ and a gluon (and viceversa),
the operators $\mathcal T_{8\,\textrm{em}}(^3P_J)$ 
destroy a singlet $Q \bar Q$ pair with quantum number $^3P_J$ and create
an octet $Q \bar Q$ pair with quantum numbers $^3S_1$ and a gluon (and viceversa).
Following the power counting  of Sec. \ref{powcou},   
the chromoelectric field scales like $(Mv)^2$ and the covariant derivative like $Mv$,
moreover, since the octet Fock-space component is suppressed by $v$, we have 
\be
\langle H(^3S_1) |  \frac{1}{M^{5}}\, \mathcal T^{(1)\prime}_{8\,\textrm{em}}(^3S_1)  | H(^3S_1) \rangle \sim M v^7,
\ee
and 
\be 
\langle H(^3P_J) |  \frac{1}{M^{5}}\, \mathcal T_{8\,\textrm{em}}(^3P_J)  | H(^3P_J) \rangle \sim M v^7.
\ee

Matrix elements involving dimension 10 octet operators are negligible at order $v^7$. 
An example is the matrix element 
\be
\langle H(^3P_0) | \frac{1}{M^{6}}\, \psi^{\dag}  \vec B  \cdot \overleftrightarrow D \chi |0\rangle 
\langle 0| \chi^{\dag} \overleftrightarrow D \cdot \vec{\sigma} \psi | H(^3P_0) \rangle  \sim M v^8.
\ee

We have kept ambiguous the ordering of the covariant derivatives appearing 
in some of the singlet operators of dimension 10. For instance,
$\overleftrightarrow D^4$ could stand either for $(\overleftrightarrow D^2)^2$
or $\overleftrightarrow D^i \overleftrightarrow D^2 \overleftrightarrow D^i$ 
or $\overleftrightarrow D^i \overleftrightarrow D^j \overleftrightarrow D^i
\overleftrightarrow D^j$ or combinations of them.
The ambiguity can only be resolved by octet operators of dimension 10, which
are beyond the present accuracy. 

All operators involved in the matching are listed in Appendix \ref{AppA}.

\subsection{Electromagnetic decay widths}
Having assumed a power counting and having chosen a basis of operators, 
we are in the position to provide explicit factorization formulas 
for S- and P-wave electromagnetic decay widths. Up to order $v^7$, these are 
\be
\label{eq:2def}
\begin{split}
&\Gamma (H(^3S_1) \rightarrow e^+ e^-) =
\frac{2\, {\rm Im} \, f_{\textrm{em}}(^3S_1)}{M^2}\langle H(^3S_1) | 
\mathcal O_{\textrm{em}} (^3S_1)| H(^3S_1)\rangle
\\ 
& + \frac{2\, {\rm Im} \, g_{\textrm{em}}(^3S_1)}{M^4}\langle H(^3S_1) | \mathcal
P_{\textrm{em}} (^3S_1)| H(^3S_1)\rangle 
+ \frac{2\, {\rm Im} \, s_{8\,\textrm{em}}(^3S_1) }{M^4} \langle H(^3S_1) | \mathcal
S_{8\,\textrm{em}}(^3S_1)| H(^3S_1)\rangle  
\\ 
& +\frac{2\, {\rm Im} \, h'_{\textrm{em}}(^3S_1)}{M^6}\langle H(^3S_1) | \mathcal
Q'_{\textrm{em}} (^3S_1)| H(^3S_1)\rangle
+  \frac{2\, {\rm Im} \, h''_{\textrm{em}}(^3S_1)}{M^6}\langle H(^3S_1) | \mathcal
Q''_{\textrm{em}} (^3S_1)| H(^3S_1)\rangle  
\\ 
& + \frac{2\, {\rm Im} \, g_{\textrm{em}}(^3S_1,^3D_1)}{M^4}\langle H(^3S_1) |
\mathcal P_{\textrm{em}} (^3S_1,^3D_1)| H(^3S_1)\rangle
\\
& 
+ \frac{2\, {\rm Im} \, t^{(1)}_{8\,\textrm{em}}(^3S_1)}{M^5}
\langle H(^3S_1) | \mathcal T^{(1)\prime}_{8\,\textrm{em}} (^3S_1)| H(^3S_1)\rangle,
\end{split}
\ee
\be
\label{eq:5def}
\begin{split}
& \Gamma(H(^1S_0) \rightarrow \gamma \gamma) =
\frac{2\, {\rm Im} \, f_{\textrm{em}}(^1S_0)}{M^2}\langle H(^1S_0) | \mathcal
O_{\textrm{em}} (^1S_0)| H(^1S_0)\rangle
\\
& + \frac{2\, {\rm Im} \, g_{\textrm{em}}(^1S_0)}{M^4}\langle H(^1S_0) | \mathcal
P_{\textrm{em}} (^1S_0)| H(^1S_0)\rangle
+ \frac{2 \, {\rm Im} \, s_{8\,\textrm{em}}(^1S_0) }{M^4} \langle H(^1S_0) | \mathcal
S_{8\,\textrm{em}}(^1S_0)| H(^1S_0) \rangle
\\ 
& +\frac{2\, {\rm Im} \, h'_{\textrm{em}}(^1S_0)}{M^6}\langle H(^1S_0) | \mathcal
Q'_{\textrm{em}} (^1S_0)| H(^1S_0)\rangle
+\frac{2\, {\rm Im} \, h''_{\textrm{em}}(^1S_0)}{M^6}\langle H(^1S_0) | \mathcal
Q''_{\textrm{em}} (^1S_0)| H(^1S_0)\rangle,
\end{split}
\ee
\be
\label{eq:6def}
\begin{split}
&\Gamma (H(^3P_{0}) \rightarrow \gamma \gamma) =
\frac{2\, {\rm Im} \, f_{\textrm{em}}(^3P_0)}{M^4}\langle H(^3P_{0}) |\mathcal
O_{\textrm{em}} (^3P_0)| H(^3P_{0}) \rangle 
\\ 
& + \frac{2\, {\rm Im} \, g_{\textrm{em}}(^3P_0)}{M^6}\langle H(^3P_{0}) |\mathcal
P_{\textrm{em}} (^3P_0)| H(^3P_{0}) \rangle
+\frac{2\, {\rm Im} \, t_{8\,\textrm{em}}(^3P_0)}{M^5}\langle H(^3P_{0}) |\mathcal
T_{8\,\textrm{em}} (^3P_0)| H(^3P_{0}) \rangle,
\end{split}
\ee
\be
\label{eq:7def}
\begin{split}
&\Gamma (H(^3P_2) \rightarrow \gamma \gamma) =
\frac{2\, {\rm Im} \, f_{\textrm{em}}(^3P_2)}{M^4}\langle H(^3P_2) |\mathcal
O_{\textrm{em}} (^3P_2)| H(^3P_2) \rangle 
\\ 
& + \frac{2\, {\rm Im} \, g_{\textrm{em}}(^3P_2)}{M^6}\langle H(^3P_2) |\mathcal
P_{\textrm{em}} (^3P_2)| H(^3P_2) \rangle
+\frac{2\, {\rm Im} \, t_{8\,\textrm{em}}(^3P_2)}{M^5}\langle H(^3P_2) |\mathcal
T_{8\,\textrm{em}} (^3P_2)| H(^3P_2) \rangle. 
\end{split}
\ee
In \eqref{eq:2def} and \eqref{eq:5def}, the first matrix element scales like
$v^3$, the second one is suppressed by $v^2$, the third one by $v^3$ and the other ones by 
$v^4$. In \eqref{eq:6def} and \eqref{eq:7def}, the first matrix element 
scales like $v^5$ and the other ones are suppressed by $v^2$. 

Some comments are in order. In Ref. \cite{Bodwin:2002hg}, the matrix elements of
the operators $\mathcal S_{8\,\textrm{em}}(^3S_1)$, $\mathcal P_{\textrm{em}} (^3S_1,^3D_1)$, 
$ \mathcal T^{(1)\prime}_{8\,\textrm{em}} (^3S_1)$ and 
$\mathcal S_{8\,\textrm{em}}(^1S_0)$ were not included in the expressions 
of $\Gamma (H(^3S_1) \rightarrow e^+ e^-)$ and $\Gamma(H(^1S_0) \rightarrow \gamma \gamma)$
The reason is that in the power counting adopted 
there, which is different from the one discussed in Sec. \ref{secpowfour}, they 
were considered to be suppressed.\footnote{We thank G. Bodwin for communications on this point.}

In Ref. \cite{Ma:2002ev}, also the matrix element 
\be
\langle H(^3P_2)| \mathcal{G}^\prime_{\em}(^3P_2) | H(^3P_2) \rangle
\ee
of the operator 
\be
\mathcal{G}^\prime_{\em}(^3P_2) =
\frac{1}{2}
\psi^{\dag}\left(-\frac{i}{2}\right)^2\overleftrightarrow D^{(i}\overleftrightarrow D^{j)}
\left(-\frac{i}{2}\overleftrightarrow D \cdot\vec{\sigma}\right) \chi|0 \rangle 
\langle 0| \chi^{\dag}\left(-\frac{i}{2}\overleftrightarrow
D^{(i}\sigma^{j)}\right)\psi + \textrm{H.c.}
\ee
was considered to contribute to Eq. \eqref{eq:7def}. In the basis of operators
given in \eqref{eq:B.5}, it can be rewritten as 
\be 
\mathcal{G}^\prime_{\em}(^3P_2) = \mathcal P_{\textrm{em}}(^3P_2,^3F_2) + \frac{2}{5}\mathcal P_{\textrm{em}}(^3P_2).
\ee
The operator $\mathcal P_{\textrm{em}}(^3P_2)$ contributes, indeed, to the
decay width of a $^3P_2$ quarkonium and its contribution has already been 
accounted for in Eq. \eqref{eq:7def}. The question is whether the operator 
$\mathcal P_{\textrm{em}}(^3P_2,^3F_2)$ contributes at relative order $v^4$ 
to the decay width of a quarkonium whose dominant Fock state has quantum
numbers $^3P_2$. In Eq. \eqref{T3PF}, we have seen that the operator acts on one of the
quarkonium states with a term proportional to ${\bf T}^{(3)}_{ijk}\sigma^k$,
where ${\bf T}^{(3)}_{ijk}$ transforms like a spherical tensor in the $L=3$ representation 
of the rotational group. The Wigner--Eckart theorem guarantees that the matrix element
$\langle 0 | {\bf T}^{(3)}_{ijk}\sigma^k |^3P_2 \rangle$ vanishes, because we cannot
generate a $L=0$ tensor by combining a $L=3$ with a $L=1$ tensor.
Explicitly, this is reflected by the fact that
\be 
\langle 0 | {\bf T}^{(3)}_{ijk}\sigma^k |^3P_2 \rangle
\sim \int d^3p \, {\rm Tr} \, \left\{
\left(p^{(i}p^{j)}\vec p \cdot \vec \sigma - \frac{2}{5} p^2
p^{(i}\sigma^{j)}\right) \sigma^n h_{^3P_2}^{nk}(\lambda) \,p^k \right\} = 0,
\ee
where $h_{^3P_2}^{ij}(\lambda)$ is a symmetric and traceless rank-2 tensor 
that represents the polarization of the $^3P_2$ state and the trace is meant
over the Pauli matrices.

\section{Matching}
\label{matching}
In this section, we calculate the imaginary parts of the matching coefficients that appear
in Eqs. \eqref{eq:2def}-\eqref{eq:7def}. The method consists in equating
(matching) the imaginary parts of scattering amplitudes in QCD and NRQCD along 
the lines of Ref. \cite{Bodwin:1994jh}.

In the QCD part of the matching, the ingoing quark and the outgoing
antiquark are represented by the Dirac
spinors $u(\vec p)$ and $v(\vec p)$ respectively, whose explicit expressions are 
\be
\label{eq:13def}
u(\vec p) = \sqrt{\frac{E_p+M}{2E_p}}
\left(
\begin{array}{c}
\xi 
\\
\frac{\vec p \cdot \vec{\sigma}}{E_p+M}\xi
\end{array}
\right), 
\qquad 
v(\vec p) = \sqrt{\frac{E_{p}+M}{2E_{p}}} 
\left(
\begin{array}{c}
\frac{\vec p \cdot \vec {\sigma}}{E_{p}+M}\eta
\\ 
\eta
\end{array}
\right),
\ee
where $E_p = \sqrt{{\vec p}^{\,2}+M^2}$, and $\xi$ and $\eta$ are Pauli spinors.
In the NRQCD part of the matching, the ingoing  quark and the outgoing
antiquark are represented by the Pauli spinors $\xi$ and $\eta$ respectively.

Since we need to match singlet and octet-singlet transition operators, 
we consider both the scattering amplitudes $Q \bar Q \rightarrow Q \bar Q $, with
two photons or an electron loop as intermediate states, and $Q \bar Q \, g \rightarrow Q
\bar Q$, with two photons or an electron loop as intermediate states
and an external gluon in the initial state. 

In the center of mass rest frame, the energy and momentum conservation imposes the following kinematical 
constraints on the scattering  $Q \bar Q \rightarrow Q \bar Q$,
\be
|\vec p| = |\vec k|, \qquad \vec p+\vec p^{\,\prime}=0, \qquad \vec k+\vec k'=0,
\ee
and on the scattering $Q \bar Q \, g \rightarrow Q \bar Q$,
\be
E_p + E_{p'}+|\vec q| = 2E_k, \qquad \vec p+\vec p^{\,\prime} +\vec q=0, \qquad \vec k+\vec k'=0,
\ee  
where $\vec p$, $\vec p^{\,\prime}$ are the ingoing and $\vec k$, $\vec k'$ the outgoing quark and antiquark 
momenta, while $\vec q$ is the momentum of the ingoing gluon, which is on mass shell. 
Note that, in the non-relativistic expansion, the gluon momentum $|\vec q|$ is
proportional to (three-momenta)$^2/M$.

The matching does not rely on any specific power counting and can be performed 
order by order in $1/M$ \cite{Manohar:1997qy}. 
We will perform the matching up to order $1/M^6$, which is the highest power in $1/M$ appearing in 
Eqs. \eqref{eq:2def}-\eqref{eq:7def}. In practice, we expand the QCD amplitude 
with respect to all external three-momenta.
In the case of the $Q \bar Q \, g \rightarrow Q \bar Q$ scattering, 
the expansion in the gluon momentum may develop infrared singularities, i.e. 
terms proportional to $1/|\vec q|$. These terms cancel in the matching. 
We will discuss the cancellation in Sec. \ref{secIR}.

Finally we note that, since the matching does not depend on the power counting
and the scattering amplitudes do not have a definite angular momentum, the matching 
determines many more coefficients than needed in Eqs. \eqref{eq:2def}-\eqref{eq:7def}.

\subsection{$Q \bar Q \to \gamma \gamma$}
The matching of the  $Q \bar Q \to \gamma \gamma$ amplitude is performed 
by equating the sum of the imaginary parts of the QCD diagrams shown in Fig. \ref{Fig:3} 
(taken by cutting the photon propagators according to $1/k^2 \to -2\pi\, i\, \delta(k^2)
\theta(k^0)$) with the sum of all the NRQCD diagrams of the type shown in Fig. \ref{Fig:3bis}.

\FIGURE[ht]{
\epsfig{file=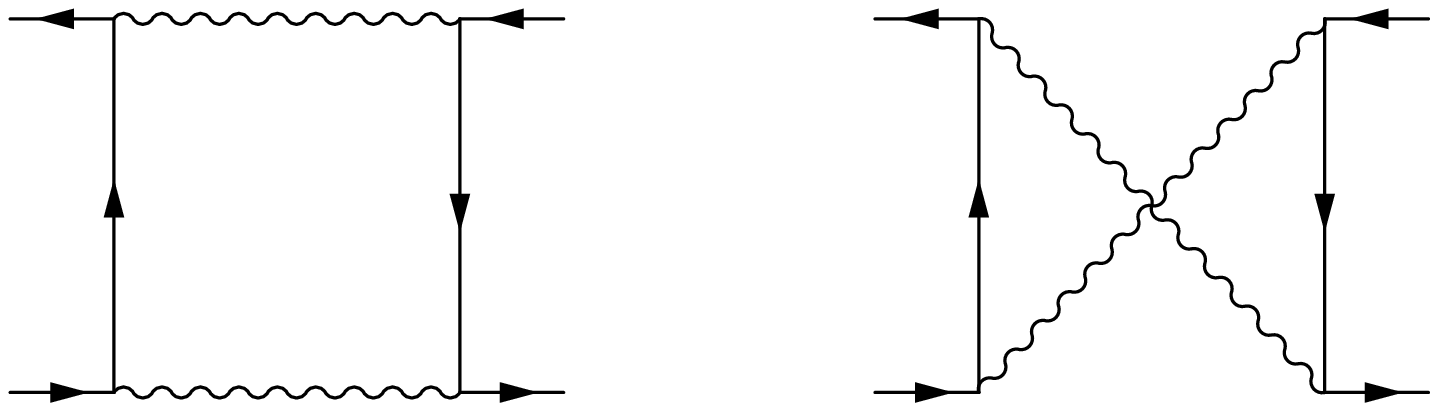,width=14cm}
\caption{QCD Feynman diagrams describing the amplitude 
$Q \bar Q \to \gamma \gamma \to Q \bar Q$ at leading order.}
\label{Fig:3}
} 

We find that 
\bea
{\rm Im} \,f_{\textrm{em}}(^1S_0) &=& \alpha^2 Q^4 \pi, 
\label{c1}
\\
{\rm Im} \,g_{\textrm{em}}(^1S_0) &=& -\frac{4}{3}\alpha^2Q^4\pi, 
\label{c2}
\\
{\rm Im} \,f_{\textrm{em}}(^3P_0) &=&  3\alpha^2Q^4\pi, 
\label{c3}
\\
{\rm Im} \,f_{\textrm{em}}(^3P_2) &=& \frac{4}{5}\alpha^2Q^4\pi, 
\label{c4}
\\
{\rm Im} \,h_{\textrm{em}}(^1D_2) &=& \frac{2}{15}\alpha^2Q^4\pi, 
\label{c5}
\\
{\rm Im} \,h'_{\textrm{em}}(^1S_0)+{\rm Im} \,h''_{\textrm{em}}(^1S_0) &=& \frac{68}{45}\alpha^2Q^4\pi,
\label{c6}
\\
{\rm Im} \,g_{\textrm{em}}(^3P_0) &=& -7\alpha^2Q^4\pi, 
\label{c7}
\\
{\rm Im} \,g_{\textrm{em}}(^3P_2) &=& -\frac{8}{5}\alpha^2Q^4\pi,  
\label{c8}
\\
{\rm Im} \,g_{\textrm{em}}(^3P_2,^3F_2) &=& -\frac{20}{21}\alpha^2Q^4\pi,
\label{c9}
\eea
where $\alpha$ is the fine structure constant and $Q$ the charge of the
quark. The four-fermion operators to which the matching coefficients 
refer are listed in Appendix \ref{AppB}.

\FIGURE[ht]{
\parbox{15cm}{
\centering
\includegraphics[width=3.5cm]{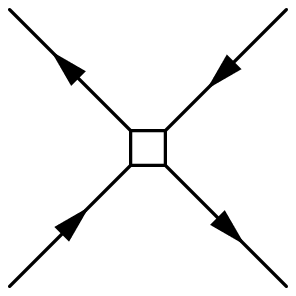}
\caption{Generic NRQCD four-fermion Feynman diagram. 
The empty box stands for one of the four-fermion vertices induced by the  
operators listed in Appendix \ref{AppA}, Eqs. \eqref{eq:B.1}-\eqref{eq:B.3}
and  \eqref{eq:B.4}-\eqref{eq:B.6}.
}
\label{Fig:3bis}
}} 

The coefficients of the operators of order $1/M^2$ and $1/M^4$ agree with
those that may be found in \cite{Bodwin:1994jh}. Here and in the
following, we also refer to \cite{Vairo:2003gh} and references therein for 
an updated list of the imaginary parts of the four-fermion matching
coefficients appearing at order $1/M^2$ and $1/M^4$ in the NRQCD Lagrangian.
Some of them are known at next-to-leading and next-to-next-to-leading order.
The coefficient ${\rm Im} \,h_{\textrm{em}}(^1D_2)$ agrees with the 
result of Ref. \cite{Novikov:1977dq}. 
Equation \eqref{c6} agrees with the one found in \cite{Bodwin:2002hg}.
By matching the diagrams of Fig. \ref{Fig:3} we cannot 
resolve ${\rm Im} \,h'_{\textrm{em}}(^1S_0)$ and ${\rm Im}
\,h''_{\textrm{em}}(^1S_0)$ separately. 
Equation \eqref{c7} agrees with the one found in \cite{Ma:2002ev}, 
whereas our evaluation of the coefficients
${\rm Im} \,g_{\textrm{em}}(^3P_2)$ and ${\rm Im} \,g_{\textrm{em}}(^3P_2,^3F_2)$
disagrees with that one in \cite{Ma:2002ev}.

\subsection{$Q \bar Q \to e^{+} e^{-}$}
The matching of the  $Q \bar Q \to e^{+} e^{-}$ amplitude is performed 
by equating the imaginary part of the QCD diagram shown in Fig. \ref{Fig:2} 
(the imaginary part of the $e^+e^-$ pair contribution to the photon's vacuum
polarization is $-\alpha\,(k^2g^{\mu\nu}-k^\mu k^\nu)/3$)
with the sum of all the NRQCD diagrams of the type shown in Fig. \ref{Fig:3bis}.

\FIGURE[ht]{
\parbox{15cm}{
\centering
\includegraphics[width=5cm]{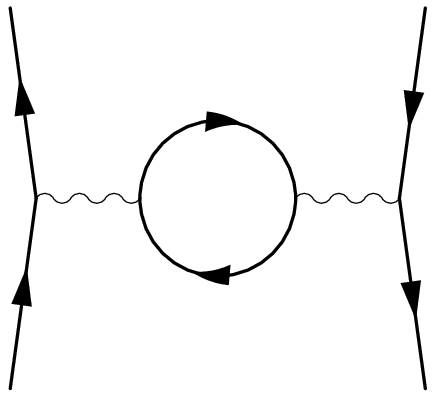} 
\caption{QCD Feynman diagram describing the amplitude 
$Q \bar Q \to e^+e^- \to Q \bar Q$ at leading order.}
\label{Fig:2}
}}

We find that
\bea
{\rm Im} \,f_{\textrm{em}}(^3S_1)&=&\frac{\alpha^2 Q^2 \pi}{3},
\label{c10}
\\
{\rm Im} \,g_{\textrm{em}}(^3S_1)&=&-\frac{4\alpha^2 Q^2\pi}{9}, 
\label{c11}
\\
{\rm Im} \,g_{\textrm{em}}(^3S_1,^3D_1)&=&-\frac{\alpha^2 Q^2 \pi}{3},
\label{c12}
\\
{\rm Im} \,h'_{\textrm{em}}(^3S_1)+{\rm Im} \,h''_{\textrm{em}}(^3S_1) &=& \frac{29}{54} \alpha^2 Q^2\pi,
\label{c13}
\\
{\rm Im} \,h_{\textrm{em}}(^3D_1) &=&\frac{\alpha^2Q^2 \pi }{12},
\label{c14}
\\
{\rm Im} \,h_{\textrm{em}}(^3D_2) &=& 0,
\label{c15}
\\
{\rm Im} \,h_{\textrm{em}}(^3D_3) &=& 0,
\label{c16}
\\
{\rm Im} \,h_{\textrm{em}}'(^3S_1,^3D_1)+{\rm Im} \,h_{\textrm{em}}''(^3S_1,^3D_1) &=& \frac{23}{36}\alpha^2Q^2\pi.
\label{c17}
\eea
The four-fermion operators to which the matching coefficients 
refer are listed in Appendix \ref{AppB}.

The coefficients of the operators of order $1/M^2$ and $1/M^4$ agree with
those that may be found in \cite{Bodwin:1994jh}. 
Equation \eqref{c13} agrees with the one found in \cite{Bodwin:2002hg} and 
Eq. \eqref{c14} with the D-wave decay width calculated in \cite{Novikov:1977dq}.
Equation \eqref{c17} is new. Equations \eqref{c15} and \eqref{c16} follow from angular momentum conservation.
Similarly to the case discussed above, we note that by matching the diagram of Fig. \ref{Fig:2}
we can only determine the sums ${\rm Im} \,h'_{\textrm{em}}(^3S_1) + {\rm Im}
\,h''_{\textrm{em}}(^3S_1)$ and ${\rm Im} \,h'_{\textrm{em}}(^3S_1,^3D_1) + {\rm Im}
\,h''_{\textrm{em}}(^3S_1,^3D_1)$ but not the individual matching coefficients.

\subsection{$Q \bar Q \, g \to \gamma \gamma$}
The matching of the  $Q \bar Q \, g \to \gamma \gamma$ amplitude is performed 
by equating the sum of the imaginary parts of the QCD diagrams shown in Fig. \ref{Fig:4}
with the sum of all the NRQCD diagrams of the type shown in Fig. \ref{Fig:4bis}.
These are all diagrams of NRQCD with an ingoing $Q \bar Q$ pair and a gluon
and an outgoing $Q \bar Q$ pair. They can involve singlet four-fermion
operators and a gluon coupled to the quark or the antiquark line, but also  
four-fermion operators that couple to gluons. Such operators induce octet
to singlet transitions on the $Q \bar Q$ pair; they may be one of the 
operators listed in Eqs. \eqref{eq:B.8} and \eqref{eq:B.9}, but also one of 
the four-fermion operators involving only covariant derivatives, which,
despite being usually denoted as singlet operators, couple to the gluon field 
through the term $-i t^a \, g \vec A^a$ in the covariant derivative.
An example is the operator 
\begin{equation*}
\mathcal{O}_{\em}(^3P_0) = 
\frac{1}{3} \psi^{\dag}
\left(-\frac{i}{2}\overleftrightarrow{D}\cdot\vec{\sigma}\right) \chi |0\rangle 
\langle 0| \chi^{\dag} \left(-\frac{i}{2}\overleftrightarrow{D}\cdot\vec{\sigma}\right)
\psi,
\end{equation*}
that annihilates (or creates) a $^3S_1$ color-octet pair and a gluon and
creates (or annihilates) a $^3P_0$ color-singlet pair through the term 
\begin{equation*}
-\frac{1}{3} \psi^{\dag}
\left(-\frac{i}{2}\overleftrightarrow{D}\cdot\vec{\sigma}\right) \chi |0\rangle 
\langle 0| \chi^{\dag} t^a \, g \vec A^a \cdot \vec{\sigma} \psi + \textrm{H.c.}.
\end{equation*}

\FIGURE[ht]{
\parbox{15cm}{
\centering
\includegraphics[width=6.5cm]{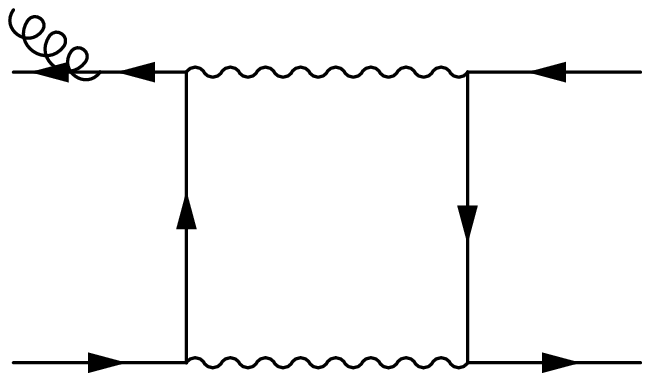}
\includegraphics[width=6.5cm]{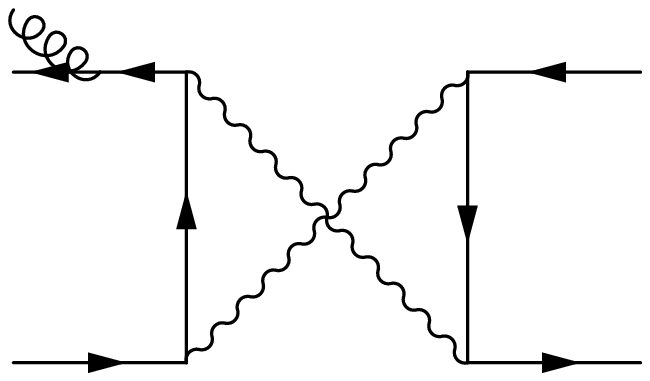}
\includegraphics[width=6.5cm]{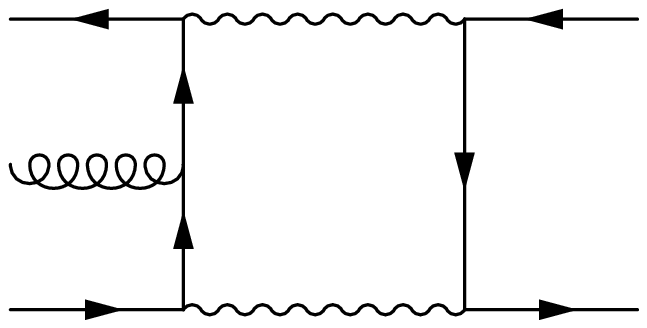}
\includegraphics[width=6.5cm]{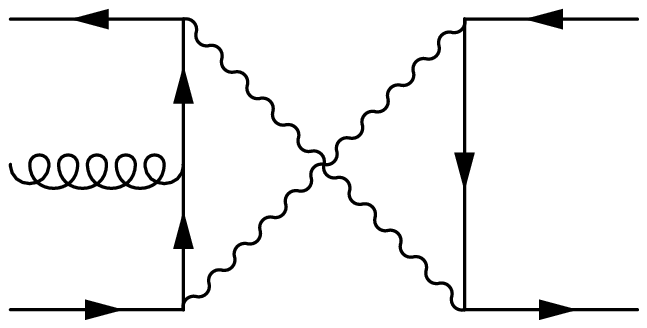}
\includegraphics[width=6.5cm]{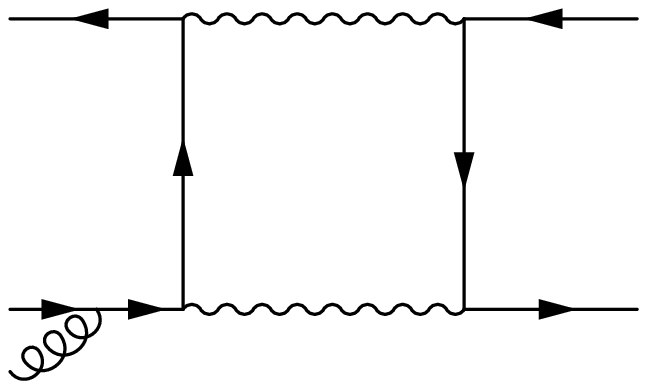}
\includegraphics[width=6.5cm]{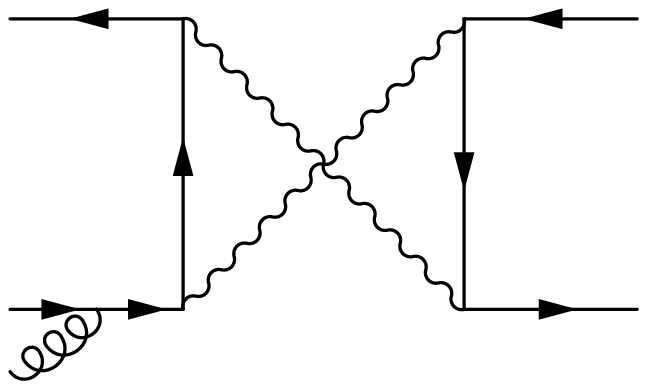}
\caption{QCD Feynman diagrams describing the amplitude 
$Q \bar Q \, g\to \gamma \gamma \to Q \bar Q$ at leading order.}
\label{Fig:4}}}

\FIGURE[ht]{
\parbox{15cm}{
\centering
\includegraphics[width=3.5cm]{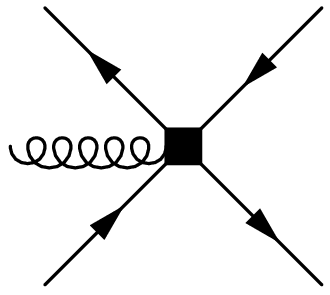}
\includegraphics[width=3.5cm]{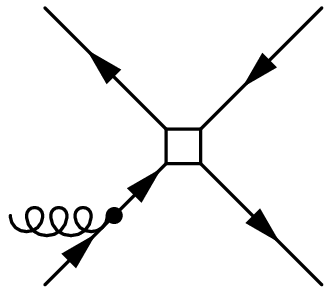}
\includegraphics[width=3.5cm]{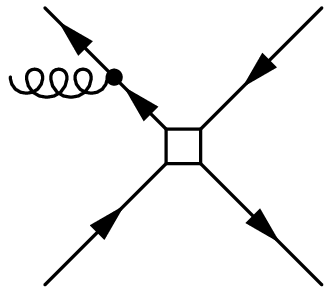}
\caption{Generic NRQCD four-fermion Feynman diagrams involving an ingoing 
$Q \bar Q$ pair and a gluon and an outgoing $Q \bar Q$ pair.
The black box with a gluon attached to it and the empty box 
stand respectively for one of the four-fermion-one-gluon vertices  
and for one of the four-fermion vertices induced by the 
operators listed in Appendix \ref{AppA}, Eqs. \eqref{eq:B.1}-\eqref{eq:B.6}.
The black dot with a gluon attached to it stands for one of the quark-gluon vertices induced by the 
bilinear part of the NRQCD Lagrangian given in Eq. \eqref{NRQCD:bilinear}.
}
\label{Fig:4bis}
}} 

In the basis of Sec. \ref{secFR}, we obtain 
\bea
{\rm Im} \,s_{8\,\textrm{em}}(^1S_0) & =& 0, 
\label{c18}
\\
{\rm Im} \,h\,'_{\textrm{em}}(^1S_0) &=& \frac{10}{9} \alpha^2 Q^4 \pi,
\label{c19}
\\
{\rm Im} \,h\,''_{\textrm{em}}(^1S_0) &=& \frac{2}{5} \alpha^2 Q^4 \pi,
\label{c20}
\\
{\rm Im} \,t_{8\,\textrm{em}}(^3P_0) &=& -\frac{3}{2}\alpha^2 Q^4\pi,
\label{c21}
\\
{\rm Im} \,t_{8\,\textrm{em}}(^3P_1) &=& 0,
\label{c22}
\\
{\rm Im} \,t_{8\,\textrm{em}}(^3P_2) &=& 0.
\label{c23}
\eea
The four-fermion operators to which the matching coefficients 
refer are listed in Appendix \ref{AppB}.

Equations \eqref{c18}-\eqref{c20} are new.
We see that the matching of the amplitude  $Q \bar Q \, g \to \gamma \gamma$ allows to 
establish the values of the individual coefficients ${\rm Im} \,h\,'_{\textrm{em}}(^1S_0)$ 
and ${\rm Im} \,h\,''_{\textrm{em}}(^1S_0)$.
Equations \eqref{c21} and \eqref{c23} disagree with those in \cite{Ma:2002ev}.
Equation \eqref{c22} follows from the Landau--Yang theorem.

\subsection{$Q \bar Q \, g \to e^+ e^-$}
The matching of the  $Q \bar Q \, g \to e^+ e^-$ amplitude is performed 
by equating the sum of the imaginary parts of the QCD diagrams shown in Fig. \ref{Fig:2.b}
with the sum of all the NRQCD diagrams of the type shown in Fig. \ref{Fig:4bis}.

In the basis of Sec. \ref{secFR}, we obtain 
\bea
{\rm Im} \,s_{8\,\textrm{em}}(^3S_1) & =& \frac{1}{3}\alpha^2 Q^2 \pi, 
\label{c25}
\\
{\rm Im} \,t^{(1)}_{8\,\textrm{em}}(^3S_1) & =& \frac{1}{6}\alpha^2 Q^2 \pi,
\label{c26}
\\
{\rm Im} \,h'_{\textrm{em}}(^3S_1) & =& \frac{23}{54}\alpha^2 Q^2 \pi,
\label{c27}
\\
{\rm Im} \,h''_{\textrm{em}}(^3S_1) & =& \frac{1}{9}\alpha^2 Q^2 \pi,
\label{c28}
\\
{\rm Im} \,h'_{\textrm{em}}(^3S_1,^3D_1) & =& \frac{5}{9}\alpha^2 Q^2 \pi,
\label{c29}
\\
{\rm Im} \,h''_{\textrm{em}}(^3S_1,^3D_1) & =& \frac{1}{12}\alpha^2 Q^2 \pi.
\label{c30}
\eea
The four-fermion operators to which the matching coefficients 
refer are listed in Appendix \ref{AppB}.

Equations \eqref{c25}-\eqref{c30} are new.
We see that the matching of the amplitude  $Q \bar Q \, g \to e^+ e^-$ allows to 
establish the values of the individual coefficients ${\rm Im}
\,h\,'_{\textrm{em}}(^3S_1)$, ${\rm Im} \,h\,''_{\textrm{em}}(^3S_1)$, 
${\rm Im} \,h\,'_{\textrm{em}}(^3S_1,^3D_1)$ and ${\rm Im}
\,h\,''_{\textrm{em}}(^3S_1,^3D_1)$, whose sums were determined by the 
$Q \bar Q \to e^+ e^-$ matching.

\FIGURE[ht]{
\parbox{15cm}{
\centering
\includegraphics[width=5cm]{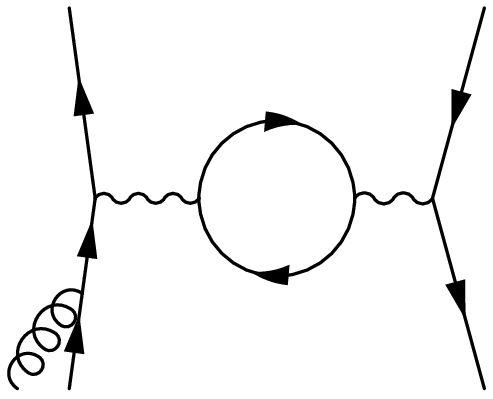}
\includegraphics[width=5cm]{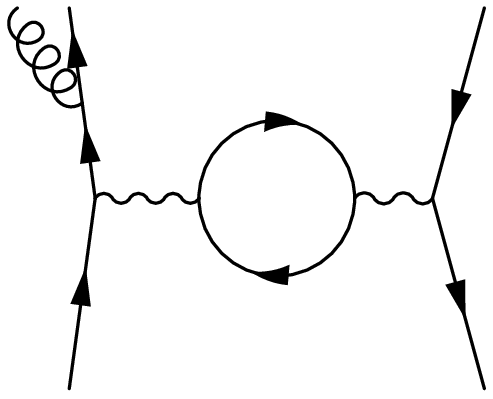}
\caption{QCD Feynman diagrams describing the amplitude 
$Q \bar Q \, g\to e^+ e^- \to Q \bar Q$ at leading order.}
\label{Fig:2.b}
}
}

\subsection{Cancellation of infrared singularities}
\label{secIR}
Once expanded in the gluon energy $|\vec q|$, diagrams with a gluon attached 
to the external quark lines generate terms that are proportional  to $1/|\vec q|$. 
They come from the almost on-shell heavy-quark propagator that follows the gluon insertion.
Such terms, which are singular for $|\vec q| \to 0$, cancel in the matching.

Let us consider two graphs like the first (second) and the fifth (sixth) in Fig. \ref{Fig:4} 
or those in Fig. \ref{Fig:2.b}. Their sum is proportional to 
\be
\bar v(\vec p^{\,\prime}) \, i t^a g \varepsilon\!\!\!/ \, 
i \frac{-\slashchar{p}' -\slashchar{q} + M}{(p'+q)^2 - M^2} \times (\ldots) 
\quad 
+
\quad 
(\ldots) \times  i \frac{\slashchar{p}+\slashchar{q}+M}{(p+q)^2 -
M^2}  \, i t^a g \varepsilon\!\!\!/ \, u(\vec p),
\label{IR1}
\ee
where $\vec\varepsilon$ is the (transverse) polarization of the external gluon.
Both quarks and gluon are on mass shell, therefore, we may write the denominator as 
\begin{equation*}
\frac{1}{(p+q)^2 - M^2} = \frac{1}{2 \left(E_p|\vec q| - \vec p \cdot \vec q \right)} .
\end{equation*}
By using the Dirac equation, we may also simplify the numerators 
\begin{equation*}
\begin{split}
\bar{v}(\vec p^{\,\prime})\,\varepsilon\!\!\!/ \, ( -\slashchar{p}' + M) =& \,
\bar{v}(\vec p^{\,\prime}) (\slashchar{p'} + M)\,\varepsilon\!\!\!/ \,  -
2 \vec p^{\,\prime}\cdot \vec \varepsilon \, \bar{v}(\vec p^{\,\prime}) 
=\,
- 2 \vec p^{\,\prime}\cdot \vec \varepsilon \, \bar{v}(\vec p^{\,\prime}) ,
\\
(\slashchar{p} + M)\,\varepsilon\!\!\!/ \, u(\vec p)  =& \,
\,\varepsilon\!\!\!/ \, (-\slashchar{p} + M ) u(\vec p) + 2 \vec p\cdot \vec
\varepsilon \,u(\vec p) = 2 \vec p \cdot \vec \varepsilon \, u(\vec p).
\end{split}
\end{equation*}
Then Eq. \eqref{IR1} becomes 
\be
\left(
- \vec p \cdot g \vec \varepsilon \frac{1}{E_p   |\vec q| - \vec p \cdot \vec q}
+ \vec p^{\,\prime}\cdot g \vec \varepsilon \frac{1}{E_{p'}|\vec q| - \vec p^{\,\prime}\cdot \vec q}
 \right) 
\, \Big(1+{O}(|\vec q|) \Big)
\times {\cal M}^a,
\label{IR2}
\ee
where ${\cal M}^a$ is the amplitude calculated from the diagrams without gluon
insertions but a color matrix $t^a$ inserted between the ingoing quark and antiquark.
Equation \eqref{IR2} displays only terms that are singular for $|\vec q| \to 0$.

In order to show the cancellation of the singular terms in the matching, 
it is sufficient to consider in NRQCD the last two diagrams of
Fig. \ref{Fig:4bis}, where the gluons couple to the quark and antiquark through 
\be
\label{eq:27def}
- \psi^{\dag} i \frac{t^a g\vec A^a}{M} \cdot \vec{\partial} \left( 1  + \frac{\vec{\partial}^{\,2} }{2M^2} +
 \frac{3}{8}\frac{\vec{\partial}^4}{M^4} + \ldots \right) \psi +  \textrm{c.c.}.
\ee
Equation \eqref{eq:27def} is the one-gluon part of ($\vec D = \vec{\partial} - i t^a \, g \vec A^a$)
\begin{equation}
\label{eq:26def}
\delta {\cal L}_{\textrm{2-f}}  = 
\psi^{\dag}\left( \frac{\vec D^2}{2M}  + \frac{(\vec D^2)^2 }{8M^3} +
 \frac{\vec D^6}{16M^5} + \ldots \right) \psi + \textrm{c.c.}.
\end{equation} 
Vertices involving $\vec E$ or $\vec B$ fields are proportional to $\vec q$. 
Hence, graphs with such vertices are not singular for $|\vec q| \to 0$.
In momentum space, the vertex induced by Eq. \eqref{eq:27def} on the quark line of Fig. \ref{Fig:4bis} is 
\begin{equation}
\label{eq:28def}
i t^a \frac{\vec{p} \cdot g \vec{\varepsilon}}{M} 
\left(1 -\frac{p^2}{2M^2} +\frac{3p^4}{8M^4} + \ldots  \right) 
= i t^a \frac{\vec{p} \cdot g \vec{\varepsilon}}{E_p};
\end{equation}
similarly, the vertex induced on the antiquark line is 
\begin{equation}
\label{eq:28defbis}
- i t^a \frac{\vec{p}^{\,\prime} \cdot g \vec{\varepsilon}}{E_{p'}}.
\end{equation}
The heavy-quark propagator in the second diagram of Fig. \ref{Fig:4bis} reads
\be
\frac{i}{E_p+|\vec q|-E_{p+q}} = \frac{i}{|\vec q| - \displaystyle \frac{\vec
p \cdot \vec q}{E_p}}
\, \Big(1+{O}(|\vec q|) \Big);
\ee
where $E_{p+q} = \sqrt{M^2 + (\vec p + \vec q)^2}$;
the heavy antiquark propagator in the third diagram of Fig. \ref{Fig:4bis} reads
\be
\frac{i}{E_{p'}+|\vec q|-E_{p'+q}} = \frac{i}{|\vec q| - \displaystyle \frac{\vec
p^{\,\prime} \cdot \vec q}{E_{p'}}}\,\Big(1+{O}(|\vec q|) \Big).
\ee
Finally, the sum of the two diagrams can be written as
\be
\left(
- \vec p \cdot g \vec \varepsilon \frac{1}{E_p   |\vec q| - \vec p \cdot \vec q}
+ \vec p^{\,\prime}\cdot g \vec \varepsilon \frac{1}{E_{p'}|\vec q| - \vec p^{\,\prime}\cdot \vec q}
 \right) \,\Big(1+{O}(|\vec q|) \Big) \times {\cal M}^a_\textrm{NRQCD},
\label{IR3}
\ee
where ${\cal M}^a_\textrm{NRQCD}$ is the NRQCD amplitude calculated from the diagrams without gluon
insertions but a color matrix $t^a$ inserted between the ingoing quark and antiquark.
Equation \eqref{IR3} displays only terms that are singular for $|\vec q| \to 0$.

Since the matching without external gluons guarantees that ${\cal M} =
{\cal M}_\textrm{NRQCD}$ and also that  ${\cal M}^a =  {\cal M}^a_\textrm{NRQCD}$,
by matching Eq. \eqref{IR2} with Eq. \eqref{IR3}, all the displayed 
singular terms cancel. In general,  the non singular terms do not cancel. 
Indeed, diagrams like the first (second) and the fifth (sixth)
of Fig. \ref{Fig:4} and those in Fig. \ref{Fig:2.b} contribute 
to the matching coefficients of NRQCD.  

We have shown by an explicit calculation how infrared singularities cancel in the
matching at leading order in $\als$ and to all orders in $1/M$. 
The argument reflects the very general fact that QCD and NRQCD 
share the same infrared behaviour.

\section{Conclusions}
\label{conclusions}
We have performed the matching of the imaginary parts of four-fermion 
operators up to dimension 10. The matching enables us to write 
the electromagnetic S- and P-wave quarkonium decay widths into two photons or an $e^+e^-$
pair up to order $v^7$.  In the power counting of Sec. \ref{secpowfour}, the
result reads
\bea
&& \hspace{-8mm}
\Gamma (H(^3S_1) \rightarrow e^+ e^-) =
\frac{2\alpha^2 Q^2 \pi}{3 M^2}\langle H(^3S_1) | 
\mathcal O_{\textrm{em}} (^3S_1)| H(^3S_1)\rangle
\nn
\\ 
&& \hspace{-3mm}
- \frac{8\alpha^2 Q^2 \pi}{9 M^4}\langle H(^3S_1) | \mathcal
P_{\textrm{em}} (^3S_1)| H(^3S_1)\rangle 
+ \frac{2\alpha^2 Q^2 \pi}{3 M^4} \langle H(^3S_1) | \mathcal
S_{8\,\textrm{em}}(^3S_1)| H(^3S_1)\rangle  
\nn
\\ 
&& \hspace{-3mm}
+\frac{23\alpha^2 Q^2 \pi}{27 M^6}\langle H(^3S_1) | \mathcal
Q'_{\textrm{em}} (^3S_1)| H(^3S_1)\rangle
+  \frac{2\alpha^2 Q^2 \pi}{9 M^6}\langle H(^3S_1) | \mathcal
Q''_{\textrm{em}} (^3S_1)| H(^3S_1)\rangle  
\nn
\\ 
&& \hspace{-3mm}
- \frac{2\alpha^2 Q^2 \pi}{3 M^4}\langle H(^3S_1) |
\mathcal P_{\textrm{em}} (^3S_1,^3D_1)| H(^3S_1)\rangle
+ \frac{\alpha^2 Q^2 \pi}{3 M^5}
\langle H(^3S_1) | \mathcal T^{(1)\prime}_{8\,\textrm{em}} (^3S_1)| H(^3S_1)\rangle,
\label{res1}
\\
&& \hspace{-8mm}
\Gamma(H(^1S_0) \rightarrow \gamma \gamma) =
\frac{2\alpha^2 Q^4 \pi}{M^2}\langle H(^1S_0) | \mathcal
O_{\textrm{em}} (^1S_0)| H(^1S_0)\rangle
\nn
\\
&& \hspace{-3mm}
- \frac{8\alpha^2 Q^4 \pi}{3 M^4}\langle H(^1S_0) | \mathcal
P_{\textrm{em}} (^1S_0)| H(^1S_0)\rangle
+\frac{20\alpha^2 Q^4 \pi}{9 M^6}\langle H(^1S_0) | \mathcal
Q'_{\textrm{em}} (^1S_0)| H(^1S_0)\rangle
\nn
\\ 
&& \hspace{-3mm}
+\frac{4\alpha^2 Q^4  \pi }{5 M^6}\langle H(^1S_0) | \mathcal
Q''_{\textrm{em}} (^1S_0)| H(^1S_0)\rangle,
\label{res2}
\\
&& \hspace{-8mm}
\Gamma (H(^3P_{0}) \rightarrow \gamma \gamma) =
\frac{6\alpha^2 Q^4 \pi}{M^4}\langle H(^3P_{0}) |\mathcal
O_{\textrm{em}} (^3P_0)| H(^3P_{0}) \rangle 
\nn
\\ 
&& \hspace{-3mm}
- \frac{14\alpha^2 Q^4 \pi}{M^6}\langle H(^3P_{0}) |\mathcal
P_{\textrm{em}} (^3P_0)| H(^3P_{0}) \rangle
-\frac{3\alpha^2 Q^4 \pi}{M^5}\langle H(^3P_{0}) |\mathcal
T_{8\,\textrm{em}} (^3P_0)| H(^3P_{0}) \rangle,
\label{res3}
\\
&& \hspace{-8mm}
\Gamma (H(^3P_2) \rightarrow \gamma \gamma) =
\frac{8\alpha^2 Q^4 \pi}{5 M^4}\langle H(^3P_2) |\mathcal
O_{\textrm{em}} (^3P_2)| H(^3P_2) \rangle 
\nn
\\ 
&& \hspace{-3mm}
- \frac{16\alpha^2 Q^4 \pi}{5 M^6}\langle H(^3P_2) |\mathcal
P_{\textrm{em}} (^3P_2)| H(^3P_2) \rangle.
\label{res4}
\eea
Equation \eqref{res1} has three extra terms with respect to the result of Ref. \cite{Bodwin:2002hg} 
(these are the terms proportional to 
$\langle H(^3S_1) | \mathcal S_{8\,\textrm{em}}(^3S_1)$ $|H(^3S_1)\rangle$, 
$\langle H(^3S_1) | \mathcal P_{\textrm{em}} (^3S_1,^3D_1)| H(^3S_1)\rangle$ 
and  $\langle H(^3S_1) |$ $\mathcal T^{(1)\prime}_{8\,\textrm{em}} (^3S_1)$ $|H(^3S_1)\rangle$), 
which are due to the different power counting.
The other terms in Eq. \eqref{res1} and Eq. \eqref{res2} agree with those in \cite{Bodwin:2002hg}, where,  
however, only the sums of the matching coefficients in front of 
$\langle H(^3S_1) |$ $\mathcal Q'_{\textrm{em}} (^3S_1)$ $| H(^3S_1)\rangle$ and 
$\langle H(^3S_1) |$ $\mathcal Q''_{\textrm{em}} (^3S_1)$ $| H(^3S_1)\rangle$,  and 
$\langle H(^1S_0) |$ $\mathcal Q'_{\textrm{em}} (^1S_0) $ $| H(^1S_0)\rangle$ and  
$\langle H(^1S_0) |$ $\mathcal Q''_{\textrm{em}} (^1S_0)$ $| H(^1S_0)\rangle$ 
were calculated. The coefficient of $\langle H(^3P_{0}) |$ $\mathcal
P_{\textrm{em}} (^3P_0)$ $| H(^3P_{0}) \rangle$ in \eqref{res3} is the same as 
in \cite{Ma:2002ev}, whereas the coefficients of $\langle H(^3P_{0}) |$ 
$\mathcal T_{8\,\textrm{em}} (^3P_0)$ $| H(^3P_{0}) \rangle$ in \eqref{res3}
and of $\langle H(^3P_2) |$ $\mathcal
P_{\textrm{em}} (^3P_2)$ $| H(^3P_2) \rangle$ in \eqref{res4} 
are different. Equation \eqref{res4} depends on two matrix elements  
less than the equivalent equation in  \cite{Ma:2002ev}.
We could only partially trace back the origin of these differences.
However, we stress that, at variance with the S-wave case, in the P-wave one, 
the differences with the previous literature cannot be reconciled by a 
different power counting.

In the case of S-wave decays, Eqs. \eqref{res1} and \eqref{res2} provide 
relativistic corrections with a few per cent accuracy in the charmonium case 
and with a few per mil accuracy in the bottomonium one. At present, the largest 
uncertainties in the decay widths of the pseudoscalars come from
next-to-next-to-leading order corrections 
to the matching coefficient of the dimension 6 operator and 
next-to-leading order corrections to the matching coefficient of the dimension 8 singlet operator, which 
are both unknown. They are about 10\% in the charmonium case and  
about 5\% in the bottomonium one (in accordance to the values 
of $\als(M)$ and $v$ listed in the introduction). In the charmonium case, 
the theoretical uncertainty is well below the experimental one.
In the case of the decay widths of S-wave vectors in lepton pairs, 
next-to-next-to-leading order corrections to the matching coefficient of the dimension 6 operator 
are known and the dominant uncertainty comes from next-to-leading order corrections 
to the matching coefficient of the dimension 8 singlet operator.  
It is about 10\% in the charmonium case and about 2\% in the bottomonium one.

In the case of P-wave decays, Eqs. \eqref{res3} and \eqref{res4} combined 
with the next-to-leading order expressions of 
${\rm Im} \,f_{\textrm{em}}(^3P_0)$ and ${\rm Im} \,f_{\textrm{em}}(^3P_2)$ 
\cite{Petrelli:1997ge} provide a consistent determination of the electromagnetic 
widths with an estimated theoretical error that is about 15\% in the
charmonium case and about 5\% in the bottomonium one. 
In \cite{Vairo:2004sr,Brambilla:2004wf}, it was observed that 
the inclusion of the next-to-leading order corrections 
to the matching coefficients alone (i.e. without relativistic corrections) brings the theoretical 
determination of the ratio of the charmonium $^3P_0$ and $^3P_2$
decay widths into two photons rather close to the experimental one.
This supports the assumption that relativistic corrections follow, indeed, 
the non-relativistic power counting in $v$ and are not anomalously large. 
The same observation, made for hadronic decay widths \cite{Vairo:2004sr,Brambilla:2004wf},  
urges a complete calculation of the relativistic corrections for P-wave hadronic 
decays of quarkonium, which is still missing. A first partial analysis 
is in \cite{Emanuele}.

Equations \eqref{res3} and \eqref{res4} depend on three and two non-perturbative 
matrix elements respectively.  Spin symmetry relates 
$\langle H(^3P_{0}) |\mathcal
P_{\textrm{em}} (^3P_0)| H(^3P_{0}) \rangle$ to 
$\langle H(^3P_{2}) |$ $\mathcal
P_{\textrm{em}} (^3P_2)$ $| H(^3P_{2}) \rangle$:
\be
\langle H(^3P_{2}) |\mathcal
P_{\textrm{em}} (^3P_2)| H(^3P_{2}) \rangle
= \langle H(^3P_{0}) |\mathcal
P_{\textrm{em}} (^3P_0)| H(^3P_{0}) \rangle \, (1+{O}(v)).
\ee
The matrix elements $\langle H(^3P_{0}) |$ $\mathcal
O_{\textrm{em}} (^3P_0)$ $| H(^3P_{0}) \rangle$ and 
$\langle H(^3P_{2}) |$ $\mathcal
O_{\textrm{em}} (^3P_2)$ $| H(^3P_{2}) \rangle$ satisfy an analogous 
relation, which, however, is not useful at relative order $v^2$. 
Therefore, Eqs. \eqref{res3} and \eqref{res4} depend on four independent 
matrix elements. Since, at present, they are poorly known, 
in this work we refrain from phenomenological applications.
We remark, however, that a further reduction in the number of non-perturbative parameters 
may be achieved in low energy EFTs like potential NRQCD (pNRQCD), where,  
in the strongly-coupled regime, the NRQCD matrix elements factorize in the
quarkonium wave function and in few correlators of gluonic fields \cite{Brambilla:2001xy}. 
The completion of the P-wave hadronic decay width calculation at order $v^7$
in NRQCD is the first step towards a complete analysis of both electromagnetic and hadronic 
relativistic corrections in the framework of the pNRQCD factorization. 
Besides direct lattice evaluations of the matrix elements
\cite{Bodwin:2005gg}, such an analysis has the  potential to constrain the non-perturbative
parameters enough to provide theoretical determinations matching 
the precision of the data.

\acknowledgments
We thank Geoffrey Bodwin and Jian-Ping Ma for correspondence and useful 
discussions and Estia Eichten for comments.
A.V. acknowledges the financial support obtained inside the Italian
MIUR program  ``incentivazione alla mobilit\`a di studiosi stranieri e
italiani residenti all'estero''. A.V. and E.M. were funded by the 
Marie Curie Reintegration Grant contract MERG-CT-2004-510967.

\appendix
\section{Summary of NRQCD operators}
\label{AppA}
The two-fermion sector of the NRQCD Lagrangian relevant for the matching discussed in Sec. \ref{matching} 
is:
\bea
{\cal L}_{\textrm{2-f}}  &=& 
\psi^\dagger \left(i D_0 + \frac{{\vec D}^2}{2M} 
+ \frac{{\vec \sigma} \cdot g {\vec B}}{2 M} 
+ \frac{({\vec D}\cdot g{\vec E})}{8 M^2}   
- \frac{{\vec \sigma} \cdot [-i {\vec D}\times, g{\vec E}]}{8 M^2}   
+ \frac{({\vec D}^2)^2}{8M^3} 
+ \frac{\{ {\vec D}^2, {\vec \sigma} \cdot g {\vec B} \}}{8 M^3}
\right.
\nn\\
&&
\left.
- \frac{3}{64 M^4} \{ {\vec D}^2, {\vec \sigma} \cdot [-i {\vec D}\times,
  g{\vec E}] \}
+ \frac{3}{64 M^4} \{ {\vec D}^2, ({\vec D}\cdot g{\vec E}) \}
+ \frac{{\vec D}^6}{16M^5}  \right) \psi
\nn\\
&& 
+ ~ \textrm{c.c.},
\label{NRQCD:bilinear}
\eea
where $\sigma^i$ are the Pauli matrices, 
$i D_0 = i \partial_0 - t^a \, g A^a_0$, $i {\vec D} = i \vec\nabla + t^a \, g {\vec A}^a$, 
$[{\vec D} \times, {\vec E}]={\vec D} \times {\vec E} - {\vec E} \times {\vec D}$,  
${E}^i = F^{i0}$ and ${B}^i = -\epsilon_{ijk}F^{jk}/2$ ($\epsilon_{123} = 1$).
We have not displayed terms of order $1/M^6$ or smaller and matching
coefficients of ${O}(\als)$ or smaller.
Equation \eqref{NRQCD:bilinear} can be obtained by matching the QCD scattering amplitude 
of a quark on a static background gluon field along the lines of Ref. \cite{Lepage:1992tx} 
(see also Ref. \cite{Das:1993jx}).

The general structure of the four-fermion sector of the NRQCD Lagrangian projected 
on the QCD vacuum is:
\be
{\cal L}^{\em}_{\textrm{4-f}}  = \sum_n \frac{c_{\em}^{(n)}}{{M^{d_n - 4}}} \mathcal O^{(n)}_{\textrm{4-f}\,\em}.
\ee 
Here, we list the operators relevant for the matching performed in Sec. \ref{matching} 
ordered by dimension ($\overleftrightarrow D = \overrightarrow D - \overleftarrow D$).
\paragraph{Operators of dimension 6}
\be
\label{eq:B.1}
\begin{split}
\mathcal{O}_{\textrm{em}}(^1S_0) = & 
\psi^{\dag} \chi |0 \rangle \langle 0|\chi^{\dag}\psi,
\\
\mathcal O_{\textrm{em}}(^3S_1)= & 
\psi^{\dag} \vec{\sigma} \chi |0 \rangle \langle 0|\chi^{\dag} \vec{\sigma} \psi.
\end{split}
\ee

\paragraph{Operators of dimension 8}
\be
\label{eq:B.2}
\begin{split}
\mathcal{P}_{\textrm{em}}(^1S_0) = &
\frac{1}{2}
\psi^{\dag}\left(-\frac{i}{2}\overleftrightarrow D \right)^2\chi |0 \rangle 
\langle 0|  \chi^{\dag}\psi
+ \textrm{H.c.},
\\
\mathcal{P}_{\textrm{em}}(^3S_1)= & 
\frac{1}{2}
\psi^{\dag} \vec{\sigma} \chi |0 \rangle 
\langle 0| \chi^{\dag} \vec{\sigma} \left(-\frac{i}{2}\overleftrightarrow D\right)^2\psi
+\textrm{H.c.},
\\
\mathcal P_{\textrm{em}}(^3S_1,^3D_1)= &
\frac{1}{2}
\psi^{\dag} \sigma^i \chi |0 \rangle 
\langle 0| \chi^{\dag} \sigma^j \left(-\frac{i}{2}\right)^2 \overleftrightarrow D^{(i} \overleftrightarrow D^{j)}\psi
+\textrm{H.c.}.
\end{split}
\ee

\be
\label{eq:B.3}
\begin{split}
\mathcal{O}_{\textrm{em}}(^1P_1) = &
\psi^{\dag}\left(-\frac{i}{2}\overleftrightarrow D \right)\chi |0 \rangle 
\langle 0| \chi^{\dag}\left(-\frac{i}{2}\overleftrightarrow D \right)\psi,
\\
\mathcal O_{\textrm{em}}(^3P_0)= & 
\frac{1}{3}  \psi^{\dag} \left(-\frac{i}{2}\overleftrightarrow{D}\cdot\vec{\sigma}\right) \chi |0 \rangle 
\langle 0| \chi^{\dag} \left(-\frac{i}{2}\overleftrightarrow{D}\cdot\vec{\sigma}\right)\psi,
\\
\mathcal O_{\textrm{em}}(^3P_1)= & 
\frac{1}{2}  \psi^{\dag}\left(-\frac{i}{2}\overleftrightarrow{D}\times\vec{\sigma}\right) \chi |0 \rangle 
\langle 0| \chi^{\dag} \left(-\frac{i}{2}\overleftrightarrow{D}\times\vec{\sigma}\right) \psi,
\\
\mathcal O_{\textrm{em}}(^3P_2)= & 
\psi^{\dag}\left(-\frac{i}{2}\overleftrightarrow{D}^{(i}\sigma^{j)}\right)\chi |0 \rangle 
\langle 0| \chi^{\dag} \left(-\frac{i}{2}\overleftrightarrow{D}^{(i}\sigma^{j)}\right) \psi.
\end{split}
\ee

\be
\label{eq:B.8}
\begin{split}
\mathcal S_{8\,\textrm{em}}(^1S_0) = &
\frac{1}{2}
\psi^{\dag} g \vec B \cdot {\vec\sigma} \chi |0 \rangle 
\langle 0| \chi^{\dag} \psi 
+ \textrm{H.c.},
\\
\mathcal S_{8\,\textrm{em}}(^3S_1) = &
\frac{1}{2}
\psi^{\dag} g \vec B \chi|0 \rangle 
\langle 0| \chi^{\dag}\vec{\sigma}\psi 
+ \textrm{H.c.}.
\end{split}
\ee

\paragraph{Operators of dimension 9}
\be
\label{eq:B.9}
\begin{split}
\mathcal T_{8\,\textrm{em}}(^1S_0) =&
\frac{1}{2} 
\psi^{\dag}\chi |0 \rangle 
\langle 0| \chi^{\dag} (\overleftrightarrow D \cdot  g\vec{E}+ g\vec E \cdot \overleftrightarrow{D} )\psi 
+ \textrm{H.c.}, 
\\
\mathcal T^{(0)}_{8\,\textrm{em}}(^3S_1) =&
\frac{1}{6}
\psi^{\dag}\vec{\sigma}\chi|0 \rangle 
\langle 0|\chi^{\dag} \vec{\sigma} (\overleftrightarrow D \cdot  g\vec{E}+ g\vec E \cdot \overleftrightarrow{D} )\psi 
+ \textrm{H.c.}, 
\\
\mathcal T^{(1)}_{8\,\textrm{em}}(^3S_1) =&
\frac{1}{4}
\psi^{\dag}\vec{\sigma}\chi|0 \rangle 
\langle 0| \chi^{\dag} \vec{\sigma} \times (-\overleftrightarrow D \times  g \vec{E} -  g \vec E \times \overleftrightarrow{D})\psi 
+ \textrm{H.c.}, 
\\
\mathcal T^{(1)\prime}_{8\,\textrm{em}}(^3S_1) =&
\frac{1}{4}
\psi^{\dag}\vec{\sigma}\chi|0 \rangle 
\langle 0| \chi^{\dag} \vec{\sigma} \times (\overleftrightarrow D \times  g \vec{E} -  g \vec E \times \overleftrightarrow{D})\psi 
+ \textrm{H.c.}, 
\\
\mathcal T^{(2)}_{8\,\textrm{em}}(^3S_1) =&
\frac{1}{2}
\psi^{\dag}\sigma^i\chi|0 \rangle 
\langle 0| \chi^{\dag} \sigma^j (\overleftrightarrow D^{(i}  g \vec{E}^{j)}+ g \vec E^{(i} \overleftrightarrow{D}^{j)})\psi
+\textrm{H.c.},
\\
\mathcal T_{8\,\textrm{em}}(^3P_0) =& 
\frac{1}{6} 
\psi^{\dag} \left(\overleftrightarrow D \cdot \vec{\sigma}\right) \chi |0 \rangle 
\langle 0|\chi^{\dag} \vec{\sigma}\cdot g \vec{E} \psi
+\textrm{H.c.},
\\
\mathcal T_{8\,\textrm{em}} (^3P_1) =& 
\frac{1}{4}
\psi^{\dag} \left(\overleftrightarrow D \times \vec{\sigma}\right) \chi |0 \rangle
\langle 0|\chi^{\dag} \vec{\sigma}\times g \vec{E} \psi
+\textrm{H.c.}, 
\\
\mathcal T_{8\,\textrm{em}} (^3P_2) =& 
\frac{1}{2}
\psi^{\dag} \left(\overleftrightarrow  D^{(i} \sigma^{j)}\right) \chi |0 \rangle 
\langle 0|\chi^{\dag} \sigma^{(i} gE^{j)} \psi 
+\textrm{H.c.}.
\end{split}
\ee

\paragraph{Operators of dimension 10}
\be
\label{eq:B.4}
\begin{split}
\mathcal{Q'}_{\textrm{em}}(^1S_0) = &
\psi^{\dag}\left(-\frac{i}{2}\overleftrightarrow D \right)^2\chi|0 
\rangle \langle 0|\chi^{\dag}\left(-\frac{i}{2}\overleftrightarrow D \right)^2\psi,
\\
\mathcal{Q''}_{\textrm{em}}(^1S_0) = &
\frac{1}{2}
\psi^{\dag}\left(-\frac{i}{2}\overleftrightarrow D \right)^4\chi  |0 \rangle 
\langle 0|\chi^{\dag}\psi
+ \textrm{H.c.},
\\
\mathcal Q'_{\textrm{em}}(^3S_1)= &
\psi^{\dag}\left(-\frac{i}{2}\overleftrightarrow D \right)^2 \vec{\sigma}\chi |0 \rangle
\langle 0| \chi^{\dag} \left(-\frac{i}{2}\overleftrightarrow D \right)^2 \vec{\sigma}\psi,
\\
\mathcal Q_{\textrm{em}}''(^3S_1)= &
\frac{1}{2}
\psi^{\dag} \left(-\frac{i}{2}\overleftrightarrow D \right)^4 \vec{\sigma} \chi |0 \rangle 
\langle 0| \chi^{\dag}\vec{\sigma}\psi 
+\textrm{H.c.},
\\
\mathcal Q_{\textrm{em}}'(^3S_1,^3D_1)= &
\frac{1}{2}
\psi^{\dag} \left(-\frac{i}{2}\right)^2 \overleftrightarrow D^{(i} \overleftrightarrow D^{j)}\sigma^i\chi|0 \rangle 
\langle 0| \chi^{\dag}\sigma^j \left(-\frac{i}{2}\overleftrightarrow D \right)^2 \psi
+\textrm{H.c.},
\\
\mathcal Q_{\textrm{em}}''(^3S_1,^3D_1)= &
\frac{1}{2}
\psi^{\dag}\left(-\frac{i}{2}\overleftrightarrow D \right)^2 
\left(-\frac{i}{2}\right)^2 \overleftrightarrow D^{(i} \overleftrightarrow D^{j)}\sigma^i\chi |0 \rangle 
\langle 0| \chi^{\dag}\sigma^j\psi
+\textrm{H.c.}.
\end{split}
\ee

\be
\label{eq:B.5}
\begin{split}
\mathcal{P}_{\textrm{em}}(^1P_1) = &
\frac{1}{2}
\psi^{\dag}\left(-\frac{i}{2}\overleftrightarrow D \right)^2\left(-\frac{i}{2} \overleftrightarrow D^i \right)\chi|0 \rangle 
\langle 0|  \chi^{\dag}\left(-\frac{i}{2} \overleftrightarrow D^i\right)\psi
+ \textrm{H.c.},
\\
\mathcal P_{\textrm{em}}(^3P_0)= & 
\frac{1}{6}
\psi^{\dag} \left( -\frac{i}{2}\overleftrightarrow D \cdot \vec{\sigma} \right) 
\left(-\frac{i}{2}\overleftrightarrow D\right)^2 \chi |0 \rangle 
\langle 0|\chi^{\dag}\left(-\frac{i}{2}\overleftrightarrow D \cdot \vec{\sigma}\right)\psi 
+ \textrm{H.c.},
\\
\mathcal P_{\textrm{em}}(^3P_1)= & 
\frac{1}{4}
\psi^{\dag} \left( -\frac{i}{2}\overleftrightarrow D \times \vec{\sigma} \right) 
\left(-\frac{i}{2}\overleftrightarrow D\right)^2 \chi |0 \rangle 
\langle 0| \chi^{\dag}\left(-\frac{i}{2}\overleftrightarrow D \times \vec{\sigma}\right)\psi 
+ \textrm{H.c.},
\\
\mathcal P_{\textrm{em}}(^3P_2)= & 
\frac{1}{2}
\psi^{\dag}\left(-\frac{i}{2}\right)\overleftrightarrow D^{(i} \sigma^{j)}
\left(- \frac{i}{2} \overleftrightarrow  D \right)^2 \chi |0 \rangle 
\langle 0| \chi^{\dag}\left(-\frac{i}{2}\overleftrightarrow D^{(i}\sigma^{j)}\right)\psi 
+ \textrm{H.c.},
\\
\mathcal P_{\textrm{em}}(^3P_2,^3F_2)= & 
\frac{1}{2}
\psi^{\dag}\left(-\frac{i}{2}\right)^2\overleftrightarrow D^{(i}\overleftrightarrow D^{j)}
\left(-\frac{i}{2}\overleftrightarrow D \cdot\vec{\sigma}\right) \chi|0 \rangle 
\langle 0| \chi^{\dag}\left(-\frac{i}{2}\overleftrightarrow D^{(i}\sigma^{j)}\right)\psi 
\\
& -
\frac{1}{5}
\psi^{\dag}\left(-\frac{i}{2}\right)\overleftrightarrow D^{(i} \sigma^{j)}
\left(- \frac{i}{2} \overleftrightarrow  D \right)^2 \chi |0 \rangle 
\langle 0| \chi^{\dag}\left(-\frac{i}{2}\overleftrightarrow D^{(i}\sigma^{j)}\right)\psi 
+ \textrm{H.c.}.
\end{split}
\ee
\be
\label{eq:B.6}
\begin{split}
\mathcal{Q}_{\textrm{em}}(^1D_2) = &
\psi^{\dag}\left(-\frac{i}{2}\right)^2\overleftrightarrow D^{(i} \overleftrightarrow D^{j)}\chi |0 \rangle 
\langle 0| \chi^{\dag}\left(-\frac{i}{2}\right)^2\overleftrightarrow D^{(i} \overleftrightarrow D^{j)}\psi,
\\
\mathcal Q_{\textrm{em}}(^3D_3)= &
\psi^{\dag}\left(-\frac{i}{2}\right)^2 \overleftrightarrow D^{((i} \overleftrightarrow D^{j)} \sigma^{l)}\chi |0 \rangle 
\langle 0|\chi^{\dag}\left(-\frac{i}{2}\right)^2 \overleftrightarrow D^{((i} \overleftrightarrow D^{j)} \sigma^{l))}\psi, 
\\ 
\mathcal Q_{\textrm{em}}(^3D_2)= &
\frac{2}{3}
\psi^{\dag}
\left(-\frac{i}{2}\right)^2 \left(\varepsilon^{ilm}\overleftrightarrow D^{(j}  \overleftrightarrow D^{l)} \sigma^{m}
+ \frac{1}{2}\varepsilon^{ijl}\overleftrightarrow D^{(m}  \overleftrightarrow D^{l)} \sigma^{m}\right) \chi|0 \rangle 
\\ 
& \times   
\langle 0| \chi^{\dag} \left(-\frac{i}{2}\right)^2 \left(\varepsilon^{inp}\overleftrightarrow D^{(j}  \overleftrightarrow D^{n)} \sigma^{p}
+ \frac{1}{2}\varepsilon^{ijn}\overleftrightarrow D^{(p}  \overleftrightarrow D^{n)} \sigma^{p}\right) \psi,
\\
\mathcal Q_{\textrm{em}} (^3D_1) = & 
\psi^{\dag}\left(-\frac{i}{2}\right)^2 \overleftrightarrow D^{(i} \overleftrightarrow D^{j)} \sigma^{i}\chi |0 \rangle 
\langle 0| \chi^{\dag}\left(-\frac{i}{2}\right)^2 \overleftrightarrow D^{(l} \overleftrightarrow D^{j)} \sigma^{l}\psi.
\end{split}
\ee

\section{Summary of matching coefficients}
\label{AppB}
In the following, we list all imaginary parts of the matching coefficients of the electromagnetic four-fermion operators 
up to dimension 10, calculated at ${O}(1)$ in the strong coupling constant 
in Sec. \ref{matching}. The matching coefficients refer to a basis of operators from where 
$\mathcal T_8(^1S_0)$, $\mathcal T_8^{(0)}(^3S_1)$, $\mathcal T_8^{(1)}(^3S_1)$ 
and $\mathcal T^{(2)}_8(^3S_1)$ have been removed by suitable field redefinitions. 
\begin{align*}
& \textrm{\bf Operator of dim. 6}                  &  &  \textrm{\bf{Matching coefficient}} &\bf{Im\,(Value)} \\
&  \mathcal O_{\textrm{em}}(^1S_0)       &  & {\rm Im} \,f_{\textrm{em}}(^1S_0)             &\alpha^2 Q^4 \pi\\
&  \mathcal O_{\textrm{em}}(^3S_1)       &  & {\rm Im} \,f_{\textrm{em}}(^3S_1)             &\frac{1}{3} \alpha^2 Q^2 \pi\\
\end{align*}
\begin{align*}
& \textrm{\bf Operator of dim. 8}                  &  &  \textrm{\bf{Matching coefficient}} & \bf{Im\,(Value)}& \\
&  \mathcal P_{\textrm{em}}(^1S_0)       &  & {\rm Im} \,g_{\textrm{em}}(^1S_0)             &  -\frac{4}{3} \alpha^2 Q^4 &\pi\\
&  \mathcal P_{\textrm{em}}(^3S_1)       &  & {\rm Im} \,g_{\textrm{em}}(^3S_1)             &  -\frac{4}{9} \alpha^2 Q^2 &\pi\\
&  \mathcal P_{\textrm{em}}(^3S_1,^3D_1) &  & {\rm Im}  \,g_{\textrm{em}}(^3S_1,^3D_1)      &  -\frac{1}{3} \alpha^2 Q^2 &\pi\\
& \\
&  \mathcal O_{\textrm{em}}(^1P_1)       &  & {\rm Im} \,f_{\textrm{em}}(^1P_1)             &  0\quad&\\
&  \mathcal O_{\textrm{em}}(^3P_0)       &  & {\rm Im} \,f_{\textrm{em}}(^3P_0)             &  3 \alpha^2 Q^4& \pi\\
&  \mathcal O_{\textrm{em}}(^3P_1)       &  & {\rm Im} \,f_{\textrm{em}}(^3P_1)             &  0\quad&\\
&  \mathcal O_{\textrm{em}}(^3P_2)       &  & {\rm Im} \,f_{\textrm{em}}(^3P_2)             &  \frac{4}{5} \alpha^2  Q^4& \pi\\
& \\
&  \mathcal S_{8\,\textrm{em}}(^1S_0)      &  & {\rm Im} \,s_{8\,\textrm{em}}(^1S_0)        & 0\quad&\\
&  \mathcal S_{8\,\textrm{em}}(^3S_1)      &  & {\rm Im} \,s_{8\,\textrm{em}}(^3S_1)        & \frac{1}{3} \alpha^2 Q^2& \pi\\
\end{align*}
\begin{align*}
& \textrm{\bf Operator of dim. 9}                  &  &  \textrm{\bf{Matching coefficient}} & \bf{Im\,(Value)}& \\
& \mathcal T^{(1)\prime}_{8\,\textrm{em}}(^3S_1) &  & {\rm Im}  \,t^{(1)}_{8\,\textrm{em}}(^3S_1)           
                                                                                            & \frac{1}{6} \alpha^2 Q^2& \pi\\
&  \mathcal T_{8\,\textrm{em}}(^3P_0)      &  & {\rm Im} \,t_{8\,\textrm{em}}(^3P_0)        & - \frac{3}{2} \alpha^2 Q^4& \pi\\
&  \mathcal T_{8\,\textrm{em}}(^3P_1)      &  & {\rm Im} \,t_{8\,\textrm{em}}(^3P_1)        & 0\quad &\\
&    \mathcal T_{8\,\textrm{em}}(^3P_2)    &  & {\rm Im} \,t_{8\,\textrm{em}}(^3P_2)        & 0\quad &\\
\end{align*}
\begin{align*}
& \textrm{\bf Operator of dim. 10}                  &  &  \textrm{\bf{Matching coefficient}} & \bf{Im\,(Value)}& \\
&\mathcal Q'_{\textrm{em}}(^1S_0)        &  & {\rm Im} \,h'_{\textrm{em}}(^1S_0)            & \frac{10}{9}\alpha^2 Q^4& \pi\\
&\mathcal Q''_{\textrm{em}}(^1S_0)       &  & {\rm Im} \,h''_{\textrm{em}}(^1S_0)           & \frac{2}{5}\alpha^2 Q^4& \pi\\
&\mathcal Q'_{\textrm{em}}(^3S_1)        &  & {\rm Im} \,h'_{\textrm{em}}(^3S_1)            & \frac{23}{54} \alpha^2 Q^2& \pi\\
& \mathcal Q''_{\textrm{em}}(^3S_1)      &  & {\rm Im} \,h''_{\textrm{em}}(^3S_1)           & \frac{1}{9} \alpha^2 Q^2& \pi\\
&  \mathcal Q'_{\textrm{em}}(^3S_1,^3D_1)&  & {\rm Im} \,h'_{\textrm{em}}(^3S_1,^3D_1)      & \frac{5}{9} \alpha^2 Q^2& \pi\\
& \mathcal Q''_{\textrm{em}}(^3S_1,^3D_1)&  & {\rm Im} \,h''_{\textrm{em}}(^3S_1,^3D_1)     & \frac{1}{12} \alpha^2 Q^2& \pi\\
& \\
\end{align*}
\begin{align*}
&\mathcal P_{\textrm{em}}(^1P_1)         &  & {\rm Im} \,g_{\textrm{em}}(^1P_1)             & 0\quad& \\
&\mathcal P_{\textrm{em}}(^3P_0)         &  & {\rm Im} \,g_{\textrm{em}}(^3P_0)             & - 7 \alpha^2 Q^4& \pi\\
& \mathcal P_{\textrm{em}}(^3P_1)        &  & {\rm Im} \,g_{\textrm{em}}(^3P_1)             & 0\quad&\\
&  \mathcal P_{\textrm{em}}(^3P_2)       &  & {\rm Im} \,g_{\textrm{em}}(^3P_2)             & -\frac{8}{5}\alpha^2 Q^4& \pi\\
&  \mathcal P_{\textrm{em}}(^3P_2,^3F_2) &  & {\rm Im} \,g_{\textrm{em}}(^3P_2,^3F_2)       & -\frac{20}{21}\alpha^2 Q^4& \pi
& \\
&\mathcal Q_{\textrm{em}}(^1D_2)         &  & {\rm Im} \,h_{\textrm{em}}(^1D_2)             & \frac{2}{15} \alpha^2 Q^4& \pi\\
&\mathcal Q_{\textrm{em}}(^3D_1)         &  & {\rm Im} \,h_{\textrm{em}}(^3D_1)             & \frac{1}{12} \alpha^2 Q^2& \pi\\
& \mathcal Q_{\textrm{em}}(^3D_2)        &  & {\rm Im} \,h_{\textrm{em}}(^3D_2)             & 0\quad&\\
&  \mathcal Q_{\textrm{em}}(^3D_3)       &  & {\rm Im} \,h_{\textrm{em}}(^3D_3)             & 0\quad&
\end{align*}


\begin{thebibliography}{999}
%\cite{Brambilla:2004wf}
\bibitem{Brambilla:2004wf}
N.~Brambilla {\it et al.},
``Heavy quarkonium physics,''
CERN-2005-005, (CERN, Geneva, 2005)
[arXiv:hep-ph/0412158].
%%CITATION = HEP-PH 0412158;%%
See also the web page of the International Quarkonium
Working Group: http://www.qwg.to.infn.it.

%\cite{Eidelman:2004wy}
\bibitem{Eidelman:2004wy}
S.~Eidelman {\it et al.}  [Particle Data Group Collaboration],
%``Review of particle physics,''
Phys.\ Lett.\ B {\bf 592}, 1 (2004).
%%CITATION = PHLTA,B592,1;%%

%\cite{Groom:2000in}
\bibitem{Groom:2000in}
D.~E.~Groom {\it et al.}  [Particle Data Group Collaboration],
%``Review of particle physics,''
Eur.\ Phys.\ J.\ C {\bf 15}, 1 (2000).
%%CITATION = EPHJA,C15,1;%%

%\cite{Bodwin:1994jh}
\bibitem{Bodwin:1994jh}
G.~T.~Bodwin, E.~Braaten and G.~P.~Lepage,
 %``Rigorous QCD analysis of inclusive annihilation and production of heavy
%quarkonium,''
Phys.\ Rev.\ D {\bf 51}, 1125 (1995) 
[Erratum-ibid.\ D {\bf 55}, 5853 (1997)]
[arXiv:hep-ph/9407339].
%%CITATION = HEP-PH 9407339;%%

%\cite{Bodwin:2002hg}
\bibitem{Bodwin:2002hg}
G.~T.~Bodwin and A.~Petrelli ,
% ``Order $v^4$ corrections to S-wave quarkonium decay,''
Phys.\ Rev.\ D {\bf 66}, 094011 (2002)
[arXiv:hep-ph/0205210].
%%CITATION = HEP-PH 0205210;%%

%\cite{Ma:2002ev}
\bibitem{Ma:2002ev}
J.~P.~Ma and Q.~Wang,
%``Corrections for two photon decays of $\chi_{c0}$ and $\chi_{c2}$ and
%color  octet contributions,''
Phys.\ Lett.\ B {\bf 537}, 233 (2002) [arXiv:hep-ph/0203082].
%%CITATION = HEP-PH 0203082;%%

\bibitem{Emanuele}
E.~Mereghetti,
``Decadimenti elettromagnetici e adronici inclusivi del quarkonio pesante'',
Diploma Thesis (Milano, 2005).

%\cite{Brambilla:2004jw}
\bibitem{Brambilla:2004jw}
N.~Brambilla, A.~Pineda, J.~Soto and A.~Vairo,
%``Effective field theories for heavy quarkonium,''
Rev.\ Mod.\ Phys.\  {\bf 77}, 1423 (2005)
[arXiv:hep-ph/0410047].
%%CITATION = HEP-PH 0410047;%%

%\cite{Kilian:1994mg}
\bibitem{Kilian:1994mg}
W.~Kilian and T.~Ohl,
%``Renormalization of heavy quark effective field theory: Quantum action
%principles and equations of motion,''
Phys.\ Rev.\ D {\bf 50}, 4649 (1994)
[arXiv:hep-ph/9404305].
%%CITATION = HEP-PH 9404305;%%

%\cite{Manohar:1997qy}
\bibitem{Manohar:1997qy}
A.~V.~Manohar,
%``The HQET/NRQCD Lagrangian to order alpha/m**3,''
Phys.\ Rev.\ D {\bf 56}, 230 (1997)
[arXiv:hep-ph/9701294].
%%CITATION = HEP-PH 9701294;%%

%\cite{Vairo:2003gh}
\bibitem{Vairo:2003gh}
A.~Vairo,
%``A theoretical review of heavy quarkonium inclusive decays,''
Mod.\ Phys.\ Lett.\ A {\bf 19}, 253 (2004) 
[arXiv:hep-ph/0311303].
%%CITATION = HEP-PH 0311303;%%

%\cite{Novikov:1977dq}
\bibitem{Novikov:1977dq}
V.~A.~Novikov, L.~B.~Okun, M.~A.~Shifman, A.~I.~Vainshtein, M.~B.~Voloshin and V.~I.~Zakharov,
%``Charmonium And Gluons: Basic Experimental Facts And Theoretical
%Introduction,''
Phys.\ Rept.\  {\bf 41}, 1 (1978).
%%CITATION = PRPLC,41,1;%%

%\cite{Petrelli:1997ge}
\bibitem{Petrelli:1997ge}
A.~Petrelli, M.~Cacciari, M.~Greco, F.~Maltoni and M.~L.~Mangano,
%``NLO production and decay of quarkonium,''
Nucl.\ Phys.\ B {\bf 514}, 245 (1998)
[arXiv:hep-ph/9707223];
%%CITATION = HEP-PH 9707223;%%
%\cite{Hagiwara:1980nv}
%\bibitem{Hagiwara:1980nv}
K.~Hagiwara, C.~B.~Kim and T.~Yoshino,
%``Hadronic Decay Rate Of Ground State Paraquarkonia In Quantum
%Chromodynamics,''
Nucl.\ Phys.\ B {\bf 177}, 461 (1981);
%%CITATION = NUPHA,B177,461;%%
%\cite{Barbieri:1979be}
%\bibitem{Barbieri:1979be}
R.~Barbieri, E.~d'Emilio, G.~Curci and E.~Remiddi,
%``Strong Radiative Corrections To Annihilations Of Quarkonia In QCD,''
Nucl.\ Phys.\ B {\bf 154}, 535 (1979).
%%CITATION = NUPHA,B154,535;%%

%\cite{Vairo:2004sr}
\bibitem{Vairo:2004sr}
A.~Vairo,
%``Open problems in heavy quarkonium physics,''
AIP Conf.\ Proc.\  {\bf 756}, 101 (2005)
[arXiv:hep-ph/0412331].
%%CITATION = HEP-PH 0412331;%%

%\cite{Brambilla:2001xy}
\bibitem{Brambilla:2001xy}
N.~Brambilla, D.~Eiras, A.~Pineda, J.~Soto and A.~Vairo,
%``New predictions for inclusive heavy-quarkonium P wave decays,''
Phys.\ Rev.\ Lett.\  {\bf 88}, 012003 (2002)
[arXiv:hep-ph/0109130];
%%CITATION = HEP-PH 0109130;%%
%\cite{Brambilla:2002nu}
%\bibitem{Brambilla:2002nu}
N.~Brambilla, D.~Eiras, A.~Pineda, J.~Soto and A.~Vairo,
%``Inclusive decays of heavy quarkonium to light particles,''
Phys.\ Rev.\ D {\bf 67}, 034018 (2003)
[arXiv:hep-ph/0208019];
%%CITATION = HEP-PH 0208019;%%
%\cite{Brambilla:2003mu}
%\bibitem{Brambilla:2003mu}
N.~Brambilla, A.~Pineda, J.~Soto and A.~Vairo,
%``The (m Lambda(QCD))**1/2 scale in heavy quarkonium,''
Phys.\ Lett.\ B {\bf 580}, 60 (2004)
[arXiv:hep-ph/0307159].
%%CITATION = HEP-PH 0307159;%%

%\cite{Bodwin:2005gg}
\bibitem{Bodwin:2005gg}
G.~T.~Bodwin, J.~Lee and D.~K.~Sinclair,
%``Spin correlations and velocity-scaling in color-octet NRQCD matrix
%elements,''
Phys.\ Rev.\ D {\bf 72}, 014009 (2005)
[arXiv:hep-lat/0503032];
%%CITATION = HEP-LAT 0503032;%%
%\cite{Bodwin:2001mk}
%\bibitem{Bodwin:2001mk}
G.~T.~Bodwin, D.~K.~Sinclair and S.~Kim,
%``Bottomonium decay matrix elements from lattice QCD with two light
%quarks,''
Phys.\ Rev.\ D {\bf 65}, 054504 (2002)
[arXiv:hep-lat/0107011];
%%CITATION = HEP-LAT 0107011;%%
%\cite{Bodwin:1996tg}
%\bibitem{Bodwin:1996tg}
%G.~T.~Bodwin, D.~K.~Sinclair and S.~Kim,
%``Quarkonium decay matrix elements from quenched lattice QCD,''
Phys.\ Rev.\ Lett.\  {\bf 77}, 2376 (1996)
[arXiv:hep-lat/9605023].
%%CITATION = HEP-LAT 9605023;%%

%\cite{Lepage:1992tx}
\bibitem{Lepage:1992tx}
G.~P.~Lepage, L.~Magnea, C.~Nakhleh, U.~Magnea and K.~Hornbostel,
%``Improved nonrelativistic QCD for heavy quark physics,''
Phys.\ Rev.\ D {\bf 46}, 4052 (1992)
[arXiv:hep-lat/9205007].
%%CITATION = HEP-LAT 9205007;%%

%\cite{Das:1993jx}
\bibitem{Das:1993jx}
A.~K.~Das,
%``On The higher order corrections to heavy quark effective theory,''
Mod.\ Phys.\ Lett.\ A {\bf 9}, 341 (1994)
[arXiv:hep-ph/9310372].
%%CITATION = HEP-PH 9310372;%%
\end{thebibliography}
\end{document}